\pdfoutput=1

\documentclass[11pt,a4paper]{article}
\usepackage{rotating}
\usepackage{times}
\usepackage{fullpage}
\usepackage{graphicx}
\usepackage{amsmath} 
\usepackage{xcolor} 
\usepackage{booktabs}
\usepackage[hidelinks,breaklinks]{hyperref}
\usepackage{natbib}

\newcommand{\light}[1]{#1} 
\newcommand{\heavy}[1]{\mathbf{#1}} 
\newcommand{\ie}{i.e.\ }

\newcommand{\ignore}[1]{}

\newcommand{\SI}{Additional File 1}
\providecommand{\keywords}[1]
{
  \small	
  \textbf{\textit{Keywords---}} #1
}

\begin{document}

\title{Testing biological network motif significance with exponential random graph models}

\author{Alex Stivala\thanks{Institute of Computational Science, Universit\`{a} della Svizzera italiana, Lugano, 6900, Switzerland. Email: \href{mailto:alexander.stivala@usi.ch}{alexander.stivala@usi.ch}
}
  \and
  Alessandro Lomi\footnotemark[1] \textsuperscript{,}\thanks{The University of Exeter Business School, Exeter, EX4 4PU, United Kingdom}
}

\maketitle

\begin{abstract}
{Analysis of the structure of biological networks often uses
  statistical tests to establish the over-representation of motifs,
  which are thought to be important building blocks of such networks,
  related to their biological functions. However, there is
  disagreement as to the statistical significance of these motifs, and
  there are potential problems with standard methods for estimating
  this significance. Exponential random graph models (ERGMs) are a
  class of statistical model that can overcome some of the
  shortcomings of commonly used methods for testing the statistical
  significance of motifs. ERGMs were first introduced into the
  bioinformatics literature over ten years ago but have had limited
  application to biological networks, possibly due to the practical
  difficulty of estimating model parameters. Advances in estimation
  algorithms now afford analysis of much larger networks in practical
  time. We illustrate the application of ERGM to both an undirected
  protein-protein interaction (PPI) network and directed gene
  regulatory networks. ERGM models indicate over-representation of
  triangles in the PPI network, and confirm results from previous
  research as to over-representation of transitive triangles
  (feed-forward loop) in an \textit{E. coli} and a yeast regulatory
  network.  We also confirm, using ERGMs, previous research showing
  that under-representation of the cyclic triangle (feedback loop) can
  be explained as a consequence of other topological features. }
\end{abstract}


\section*{Introduction}

Molecular interactions in biological systems are often represented as
networks \citep{winterbach13}. Some such networks are inherently
undirected, such as protein-protein interaction (PPI) networks
\citep{delasrivas10}. Others may be directed, such as gene regulatory
networks, where nodes represent operons, and arcs (directed edges)
represent transcriptional interactions between them. Much research
with such biological networks has concerned ``motifs'', small
subgraphs which occur more frequently than would be expected by
chance.  Motifs have been considered the building blocks of complex
networks \citep{milo02,shen02,alon07,ciriello08}. The biological
significance of network motifs derives from their possible
interpretation as signs of evolutionary events
\citep{middendorf05,rice05}.

Two simple examples of motifs in undirected networks are
triangles (three-cycles) and squares (four-cycles) \citep{rice05}. Directed networks allow for a
larger set of potentially important motifs
\citep{milo02,middendorf05,rice05}, which can be quite complicated,
leading to problems of consistency in their definition
\citep{konagurthu08}.

\begin{figure}
  \centering
  \includegraphics[width=\textwidth]{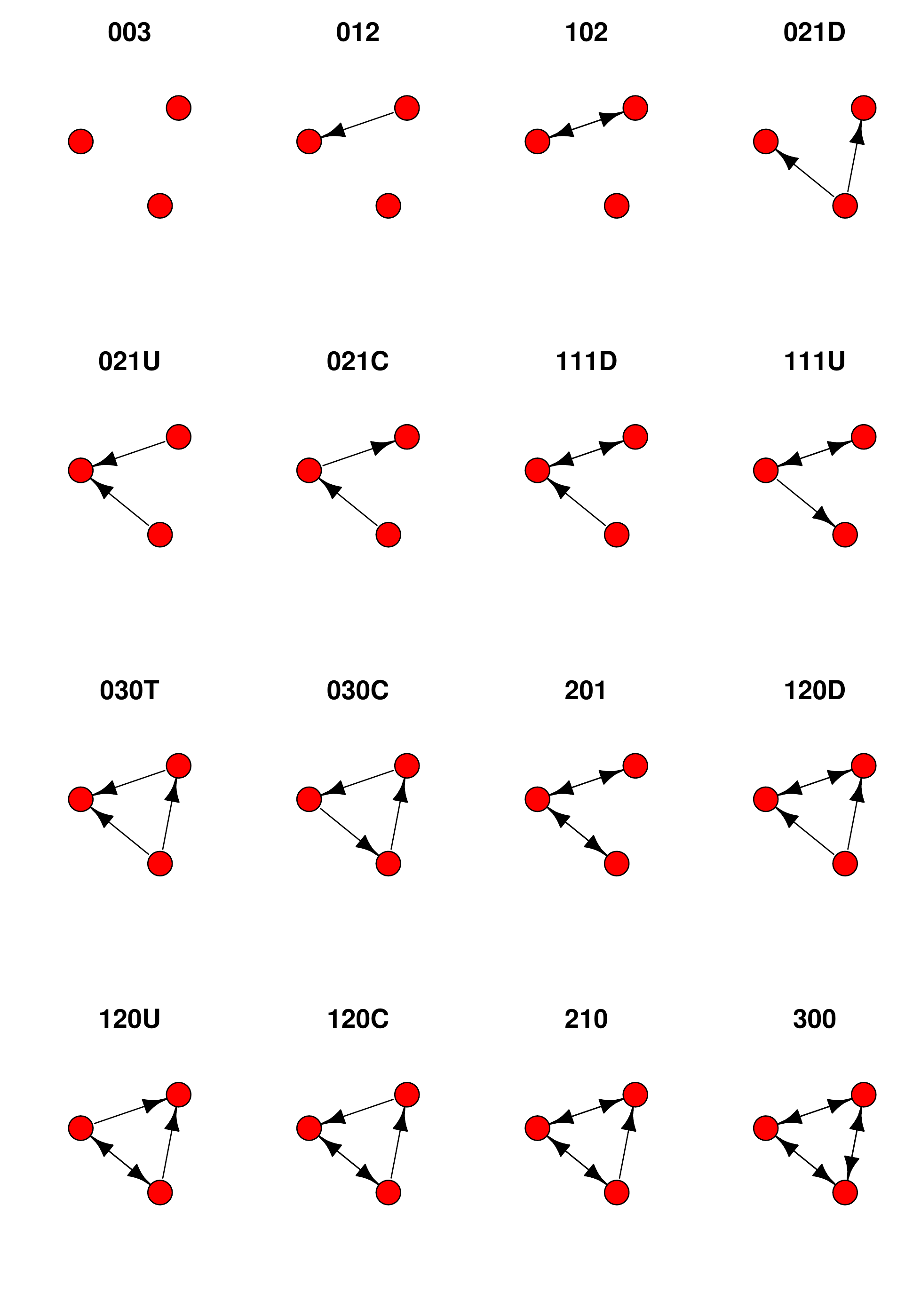}
  \caption{Triad census classes labeled with the MAN
      (mutual, asymmetric, null) dyad census naming convention.
    When the dyad census does not uniquely identify a triad, a
    letter designating ``up'', ``down'', ``transitive'', or
    ``cyclic'' is appended
    \label{fig:triad_census}}
\end{figure}

It is worth noting that such (three-node) motifs are an idea with a
long history in social network analysis, where the counts of all
sixteen possible three-node directed graphs (triads) are known as the
triad census \citep{davis67,holland70,holland76,wasserman94}. A
systematic naming convention has been developed that is based on the
number of mutual, asymmetric, and null ($M$, $A$, and $N$) dyads in
the triad, followed by a letter to distinguish the orientation if it
is not unique (Fig.~\ref{fig:triad_census}). For example, the
transitive triangle is designated 030T, which distinguishes it from
the cyclic triad 030C. Although in common usage in social network
research, and cited by \citet{milo02} and \citet{saul07} in the context of biological networks, this
naming convention is rarely used in discussions of motifs in the
bioinformatics or biology literature. There are efficient algorithms for
computing the triad census \citep{moody98,batagelj01}, implemented in
widely used general purpose graph libraries such as igraph
\citep{csardi06} and NetworkX \citep{hagberg08}. The triad census has
recently been extended to colored triads, that is, distinguishing the
nodes in the triads based on a categorical attribute assigned to them
\citep{lienert19}. It has long been noted in the social networks
literature that the dyad census constrains the triad census, and yet
empirical social networks often still have counts for some triads
greater than expected given those constraints \citep{faust10}.

To determine if a motif is over-represented, the count of the
motif in an observed network is compared to the distribution of its
counts in a set of simulated random networks \citep{ciriello08} (it
is also possible to determine the significance of motif
over-representation without simulation \citep{picard08,martorana20}). This leads to
the problem of choosing the appropriate random networks (null model),
and some supposed motifs have been found to not be significantly
over-represented, and occur with the observed frequencies simply due
to topological properties of random networks \citep{konagurthu08a} or
correlations between motifs created by the randomization process
\citep{ginoza10}, although such correlations can also occur even with
uniform sampling \citep{fodor20}.

Estimating motif (triad census) significance by comparing
the triad census of an empirical network to that of ensembles of
random graphs also has a long history, for example the conditional
uniform graph (CUG) distribution \citep{mayhew84,anderson99,butts08},
conditional on the dyad census (U\textbar MAN) \citep{holland76}, or
on the degree distribution \citep{snijders91}. A more modern variation
on a similar idea is the $dk$-series \citep{mahadevan06,orsini15}, a
sequence of nested network distributions of increasing complexity,
fitting in turn density, degree distribution, degree homophily, average local
clustering, and clustering by degree \citep{orsini15}.

The recent work of \citet{fodor20} shows that the
assumptions of mainstream methods for motif identification, specifically
normally distributed motif frequencies and independence of motifs, do
not always hold, and that, as a consequence, such methods cannot always
correctly estimate the statistical significance of motif
over-representation.

Aside from such intrinsic statistical limitations, it may be the case
that the apparent statistical over-representation of motifs has no
evolutionary or functional significance
\citep{mazurie05,ingram06,payne15}, and the choice of null model is a
critical factor in this lack of evident relationship between
over-representation and evolutionary preservation
\citep{mazurie05,beber12}. Alternatively, the apparent lack of
functional significance \citep{payne15} may be due to too narrow a
definition of ``function'' \citep{ahnert16}.  Recently, it has also
been suggested that elementary motifs are a lower level of structure
than that which is most functionally relevant in gene regulatory networks
characterizing different physiological states \citep{lesk20}.

It might also be the case that particular motifs are over-represented,
not because they are evolutionarily selected for function, but because
of spatial clustering \citep{artzy04}. For example, in the context of
PPI networks, we might expect that interactions would be
over-represented between proteins that share a subcellular location,
and under-represented between those that do not, since proteins known
to interact usually have the same subcellular locations
\citep{vonmering02}. Indeed PPI networks can be used as predictors of
subcellular location \citep{shin09,kumar10}.

There are many algorithms for motif discovery in complex networks; for
recent reviews, see \citet{patra20,jazayeri20,yu20}. In the present work
we are considering only static, not temporal, networks. Although they
differ in many details, especially regarding computational efficiency
and scalability, these motif discovery algorithms work fundamentally
in the manner described above. That is, they count occurrences of a
motif in the observed network, and compare this to the distribution of
the motif's frequency in an ensemble of randomized versions of the
original network (typically preserving degree sequence). Therefore
these conventional methods all test the significance of one motif at a
time, assuming independence of motifs, and are all potentially subject
to the problems described by the recent work of \citet{fodor20},
mentioned above.  That is, that the assumptions of independence and
normal distribution of motif frequencies may not hold, and that
therefore these methods might not be able to correctly estimate the
statistical significance of motif over-representation.

In this work we describe a different approach to determining motif
significance in complex networks, which can potentially overcome these
problems. Rather than comparing the observed frequency of a candidate motif to
its frequency in a set of randomized networks, we take a model-based
approach. Specifically, we estimate parameters of a model (an
exponential random graph model, abbreviated ERGM) of the observed
network. These parameters correspond to substructures which
resemble potential motifs of interest. This allows the significance of
the candidate motifs to be tested simultaneously in a single model, in
such a way that independence of the motifs is not assumed.

Once such a model is estimated, it can also be used to test for motif
significance in the traditional way, using the ERGM to simulate
an ensemble of random networks. Recently, this approach was used
test for motifs (dyads, triads, and tetrads; that is, two, three, and
four node motifs) in a collection of social (rather than biological)
networks \citep{felmlee21}. Using ERGM rather than
degree-preserving randomization, ``reduces the scope for misleading
results by controlling for multiple, potential correlates in the same
set of random models.''  \citep[p.~2]{felmlee21}.

We demonstrate the ERGM approach in biological networks (both undirected
(PPI) and directed gene regulatory networks) using some recently
developed ERGM estimation methods
\citep{byshkin16,byshkin18,borisenko19,stivala20}, which allow
estimation of models for larger networks than was practical with 
earlier methods of ERGM parameter estimation.

The remainder of this article is organized as follows. First, we
describe ERGMs, and review the literature on the application of
ERGMs to biological networks. We then report the biological networks
considered in this work, and the details of the ERGM configurations,
estimation methods, and goodness-of-fit tests we used. Following that, we present
and discuss new ERGM models of these networks, comparing the
inferences as to motif significance with existing published results
using conventional motif discovery methods. In the next section, we detail the
limitations of this application of ERGMs, and indicate some potential
future work. We conclude with a summary of the inferences drawn from
the ERGM models of the networks considered.

\section*{Exponential random graph models}

ERGMs are widely used in the social sciences, typically to model
social networks \citep{robins07intro,lusher13,amati18,koskinen20}.
\citet{cimini19} is a recent review of ERGMs for modeling
real-world networks, from a statistical physics viewpoint.

An ERGM is a probability distribution with the form
\begin{equation}
  \Pr(X =x) = \frac{1}{\kappa(\theta)}\exp\left(\sum_A \theta_A z_A(x)\right)
\end{equation}
where
\begin{itemize}
\item $X = [X_{ij}]$ is a 0-1 matrix of random tie variables,
\item $x$ is a realization of $X$,
\item $A$ is a ``configuration'', a (small) set of nodes and a subset of ties between them,
\item $z_A(x)$ is the network statistic for configuration $A$,
\item $\theta_A$ is a model parameter corresponding to configuration $A$,
\item $\kappa(\theta)$ is a normalizing constant to ensure a proper distribution.
\end{itemize}
Given an observed network $x$, we aim to find the parameter vector
$\theta$ which maximizes the probability of $x$ under the model. Then
for each configuration $A$ in the model, its corresponding
parameter $\theta_A$ and its estimated standard error allow us to make
inferences about the over- or under-representation of that configuration in
the observed network. If $\theta_A$ is significantly different from
zero, then if $\theta_A > 0$ the configuration $A$ is over-represented, or
under-represented if $\theta_A < 0$.

\begin{figure}
  \centering
  \includegraphics[width=0.6\textwidth]{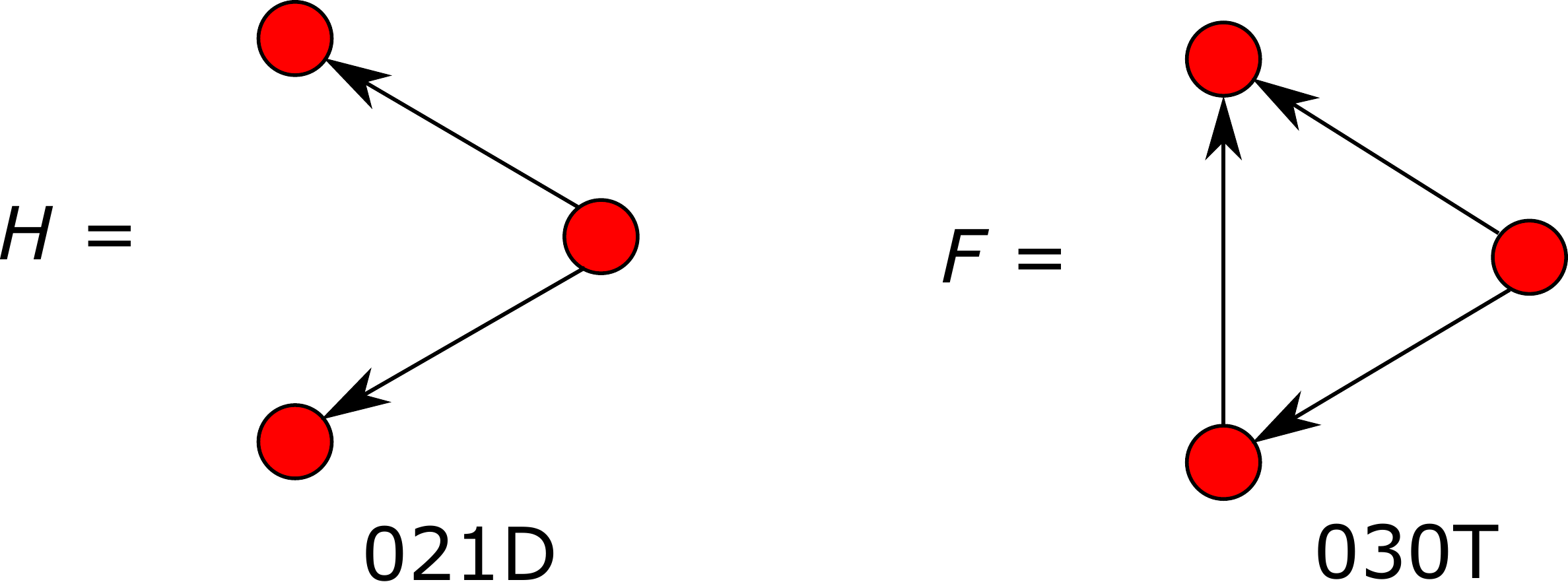}
  \caption{Motif examples.
    $F$, the transitive triangle (triad 030T) is not
    a special case of $H$, the out-star (triad 021D), when considered as
    motifs (or triad census classes): they are distinct
    induced subgraphs of three nodes. However, when considered as ERGM
    configurations, since $H$ is a subgraph (but not an induced subgraph) of $F$
    (the transitive triangle is formed by ``closing'' the out-star with an
      additional arc), in their corresponding statistics both $F$ and
      $H$ are counted for an occurrence of $F$
      \label{fig:triad_examples}}
\end{figure}

Note that a ``configuration'', unlike a motif (in its most common
usage) or the triad census classes, is not an induced
subgraph. That is, it does not include every edge in the original
graph of which it is a subgraph: a configuration is any occurrence of
the substructure in question in the graph; it is defined only by its
edges, not by its edges and non-edges. See
Fig.~\ref{fig:triad_examples} for an example based on one from 
\citet[Fig.~5B]{fodor20}.

ERGMs solve the problem of the need to correct for correlations
between motif occurrences, and also other attributes such as
subcellular location (functional and evolutionary significance is
another matter entirely). Given an observed network, model parameters
can be estimated  by maximum likelihood. Hence parameters
corresponding to candidate motifs such as triangles can be estimated, and a
positive significant parameter would indicate triangles occurring more
frequently than by chance, \emph{given the other parameters in the
  model} (which would include parameters to control for density and
degree distribution, for example). ERGMs allow different structural configurations to
be incorporated, as well as configurations based on node attributes (such as
physico-chemical properties, or spatial locality), and the
significance of the configurations can then be assessed given all the other
structural and other configurations included in the model.

ERGMs fulfill all of the desirable criteria for improved network models
listed by \citet[p.~427]{desilva05}. They take into account that
networks are finite. Indeed, far from requiring very large networks to
fit the requirements of mean-field theories, they are dependent on
network size and do not scale consistently to infinity
\citep{rolls13,shalizi13,schweinberger19} --- a property that can be
used to estimate population size from network samples \citep{rolls17}.
They can handle modular organization or community or block structure \citep{fronczak13,schweinberger15,schweinberger18,wang17c,babkin20,schweinberger20,gross21}, samples from larger networks
\citep{handcock10,pattison13,stivala16,an16}, and missing data
\citep{robins04,koskinen13a}. And finally, they are flexible at
incorporating additional information such as nodal attributes,
including dyadic attributes, such as distances between nodes. ERGMs
have also been extended to handle valued networks \citep{desmarais12,krivitsky12}
and dynamic (time-varying) networks \citep{krivitsky14}, and to
use graphlets \citep{przulj07} as the ERGM configurations \citep{yaveroglu15}.

Despite these potential advantages, however, ERGM parameter estimation is a computationally
intractable problem, and in practice it is generally necessary to use
Markov chain Monte Carlo (MCMC) methods \citep{hunter12}. A variety of
algorithms for ERGM model fitting
\citep{snijders02,hunter06,hummel12,krivitsky17} are implemented in
widely used software packages such as statnet
\citep{handcock08,morris08,hunter2008ergm} and PNet/MPNet
\citep{wang09}, and Bayesian methods are also available
\citep{caimo11,caimo14}. These packages also implement the so-called
``alternating'' or ``geometrically weighted'' configurations
\citep{snijders06,robins07}, which alleviate problems with model
``near-degeneracy'', where the model's probability mass is concentrated in
a very small region of possible networks, which can occur when only
simple configurations, such as stars and triangles, are used
\citep{hunter12}.

Until recently, the computational difficulty of ERGM parameter
estimation has limited its application to biological networks,
which are often larger than the social networks (traditionally
measured by observations and surveys, rather than online social
networks) for which the techniques were developed. Now, however,
advances such as snowball sampling and conditional estimation
\citep{pattison13,stivala16}, improved ERGM distribution samplers such
as the ``improved fixed density'' (IFD) sampler \citep{byshkin16}, and
new estimation algorithms \citep{hummel12}, including the
``Equilibrium Expectation'' (EE) algorithm
\citep{byshkin18,borisenko19} and its implementation for large
directed networks \citep{stivala20}, have reduced by orders of
magnitude the time taken to estimate ERGM parameters.

\section*{Literature review of application of ERGMs to biological networks}

ERGMs were first applied to biological networks by
\citet{saul07}, who estimated model parameters for \textit{Escherichia
  coli} \citep{salgado01} and yeast regulatory networks, and a
collection of metabolic networks. As well as introducing the use of
ERGMs to the field of bioinformatics for analyzing biological
networks, \citet{saul07} used ERGM models to build
topological profiles which they showed to be capable of classifying
organisms into biological and functional groups.  With the algorithms
and implementations available at the time, the larger networks could
only be estimated by maximum pseudo-likelihood \citep{strauss90}, an
approximation which is now considered problematic
\citep{robins07,vanduijn09,hunter12} and useful mostly for obtaining
initial parameter estimates for a more accurate (but also more
computationally expensive) method
\citep{hunter06,hummel12,krivitsky17}. Further, all the networks in
\citet{saul07} were treated as undirected, thereby
losing important directional information (and not, for example,
being able to distinguish between cyclic and transitive triads) in
regulatory networks. 
The \textit{E. coli} regulatory network, treated as undirected, was
also used as an example application of the ``stepping'' algorithm
for ERGM estimation by \citet{hummel12}.

Exponential random graph models for similar \textit{E. coli}
regulatory networks were described by \citet{begum14},
leaving the networks directed rather than treating them as
undirected. These models were very simple, however, including only Arc
and In-star terms, and therefore model degree distribution, but not
triangular motifs.

Bayesian estimation of an ERGM model of a human PPI network with 401
proteins was described by \citet{bulashevska10}.
This model used only very basic structural features (not including any
triangular structures, for example), but made use of nodal attributes,
specifically a binary variable indicating if the protein is
disordered. This ERGM was not used to analyze network motifs, but
rather the relationship between disordered proteins and their
``sociality'', a measure of their importance in the PPI network,
finding that intrinsically disordered proteins tend to be more
``social'' \citep{bulashevska10}. In their Conclusions,
\citet{bulashevska10} suggest that ``The ERGM modelling of
networks offers a natural way of assessing importance of the network
motifs'' \citep[p.~13]{bulashevska10}.

Similar techniques, that is, Bayesian estimation of ERGMs with only
very simple structural terms, have also been used with gene-gene
relationship networks to model mechanisms of gene
dysregulation \citep{azad17}. These models were used to infer potential
aberrant gene pairs, and suggested a novel pattern of aberrant
signaling \citep{azad17}.

A mixture ERGM was introduced by \citet{wang17c} and applied
to a yeast gene interaction network with 424 genes
\citep{schuldiner05,wang17c}. The model included geometrically weighted in-degree
and out-degree terms, but not any triangular terms; the interest is
rather in the clusters it finds, which may be used to predict
function \citep{wang17c}.

An ERGM incorporating a directed form of the degree-corrected
stochastic blockmodel \citep{karrer11} was introduced by
\citet{gross21}, and applied to the connectome of
the \textit{C.~elegans} worm (279 nodes representing neurons), and an
\textit{A.~thaliana} PPI network (4344 nodes representing proteins).
These models assume dyadic independence, and hence triangular
configurations could not be incorporated.  The advantage of the
mixture ERGM \citep{wang17c} or stochastic blockmodel ERGM
generalizations ($\beta$-SBM and $p_1$-SBM \citep{gross21}) is that they
can capture heterogeneity in clusters found in the network, but we do
not address cluster or community structure here.

ERGMs have been applied to neural networks with 90 nodes, representing
brain regions \citep{simpson11,simpson12}, finding that an ERGM
approach outperforms conventional approaches for constructing
group-based representative brain networks \citep{simpson12}.  Bayesian
ERGM techniques, with 96 nodes representing brain regions, have been
used to model brain networks over the human
lifespan \citep{sinke16}. Recently, Bayesian ERGMs, extended to
multiple networks, were used to compare functional connectivity
structure across groups of individuals \citep{lehmann21}.

ERGMs have also been used to model human brain
networks inferred from electroencephalographic (EEG) signals; these
networks have 56 (the number of EEG sensors) nodes \citep{obando17}.
These models showed that clustering and node centrality (as reflected
by over-representation of triangles and stars) better explained global
properties of the brain networks than other graph metrics, supporting
the view that segregated modules exchange information via hubs.

An enhanced version of the generalized (or valued) ERGM
\citep{desmarais12} was used to model the human Default Mode Network
(DMN) with 20 nodes, representing brain regions
\citep{stillman17}. This model showed that the DMN appears to be organized
in a ``segregated highway'' structure, that is, with fewer hubs and
more triadic closure than expected, in contrast to ``small world''
structure of the whole-brain network \citep{stillman17}. This work is
an example of an ERGM that incorporates spatial distances, in the form
of three-dimensional Euclidean distances between nodes.

A Bayesian ERGM has been used to model transient structure in
intrinsically disordered proteins, providing a means for identifying
transient structures that differ in favorability across
variants \citep{grazioli19}.  A specific family of ERGMs has been used
to model amyloid fibril topologies, leading to the construction of a
systemic nomenclature that can classify all known amyloid
fibril structures, and a simulation technique that can explore the
kinetics of fibril self-assembly \citep{grazioli19b}.

Simple ERGMs for undirected networks (\textit{A.~thaliana}, yeast,
human, and \textit{C.~elegans} PPI networks, and undirected versions
of \textit{E.~coli} regulatory and \textit{Drosophila} optic medulla
networks) were estimated in \citet[S.I.]{byshkin18}, demonstrating
that the EE algorithm could be used to estimate in minutes a model
that takes many hours or is practically impossible with earlier
methods. In addition, a more complex model of the \textit{A.~thaliana}
PPI network was estimated, showing not just the over-representation of
the triangle motif, but also the tendency for plant-specific proteins
to interact preferentially with each other, and for kinases to
interact preferentially with phosphorylated proteins \citep{byshkin18}.
However that work dealt only with undirected networks. An
implementation of the EE algorithm for directed networks was described
in \citet{stivala20}, but no biological networks were considered in
that work.

\section*{Methods}

\subsection*{Network data}

We obtained a yeast PPI network \citep{vonmering02} from the igraph \citep{csardi06}
Nexus network repository (this is no longer
available, we used the network downloaded on 10 November
2016).
The yeast PPI network has the proteins annotated with one of 12
functional categories \citep{mewes02,ruepp04} (or
``uncharacterized''), as described in the Supplementary Information of
\citet{vonmering02}.

We obtained a human PPI network from the HIPPIE database
\citep{schaefer12,schaefer13,suratanee14,alanislobato17}, version 2.2,
downloaded from \url{http://cbdm.uni-mainz.de/hippie/} (accessed 12
June 2021). Edges in this network are labeled with a confidence score
between zero and one.  We built a binary ``high confidence'' network
by selecting edges where the score is $\geq 0.70$, the third quartile of
the score distribution.

To annotate nodes in the human PPI network with their subcellular
location using terms in the Gene Ontology (GO) \citep{ashburner00}, we
used the PANTHER (Protein ANalysis THrough Evolutionary Relationships)
database \citep{mi21,mi19}. We used the PANTHER database version 16.0
downloaded from
\url{http://data.pantherdb.org/ftp/sequence_classifications/current_release/PANTHER_Sequence_Classification_files/PTHR16.0_human}
(accessed 21 June 2021). We used the R package GOxploreR
\citep{manjang20,goxplorer} to rank the GO terms for subcellular
component in the PANTHER database, and annotated each node
(representing a protein) in the network with the highest ranking term
for that protein.  This results in a cellular component GO term for
6~131 of the 11~517 nodes (53\%) in the human PPI network. The
cellular component GO terms are treated as a categorical attribute, of
which there are 271 unique values in the data. The nodes with no
cellular component GO term assigned are given an ``NA'' category,
which, when used in the ``Match'' statistic in ERGM modeling, does not
match any category (including the NA category itself).

The previously mentioned \textit{E. coli} regulatory network
\citep{salgado01,shen02} was obtained via the statnet package
\citep{handcock08,statnet}. Following \citet{hummel12}, we removed the
loops (self-edges) representing self-regulation, and considered
self-regulation instead in a simplistic way by a binary node attribute
designated ``self'' which is true when a self-loop was present and
false otherwise. In some models, we use the original version of this
network with self-edges retained, and when this is done it is noted in
the results.  We also obtained a \textit{Saccharomyces cerevisiae}
(yeast) regulatory network \citep{milo02,costanzo01}
(\url{http://www.weizmann.ac.il/mcb/UriAlon/download/collection-complex-networks};
accessed 29 April 2019) and processed it in the same way.

For all networks, we removed multiple edges and, unless noted otherwise, self-loops, where these are present.

\begin{figure}
  \centering
 \includegraphics[width=\textwidth]{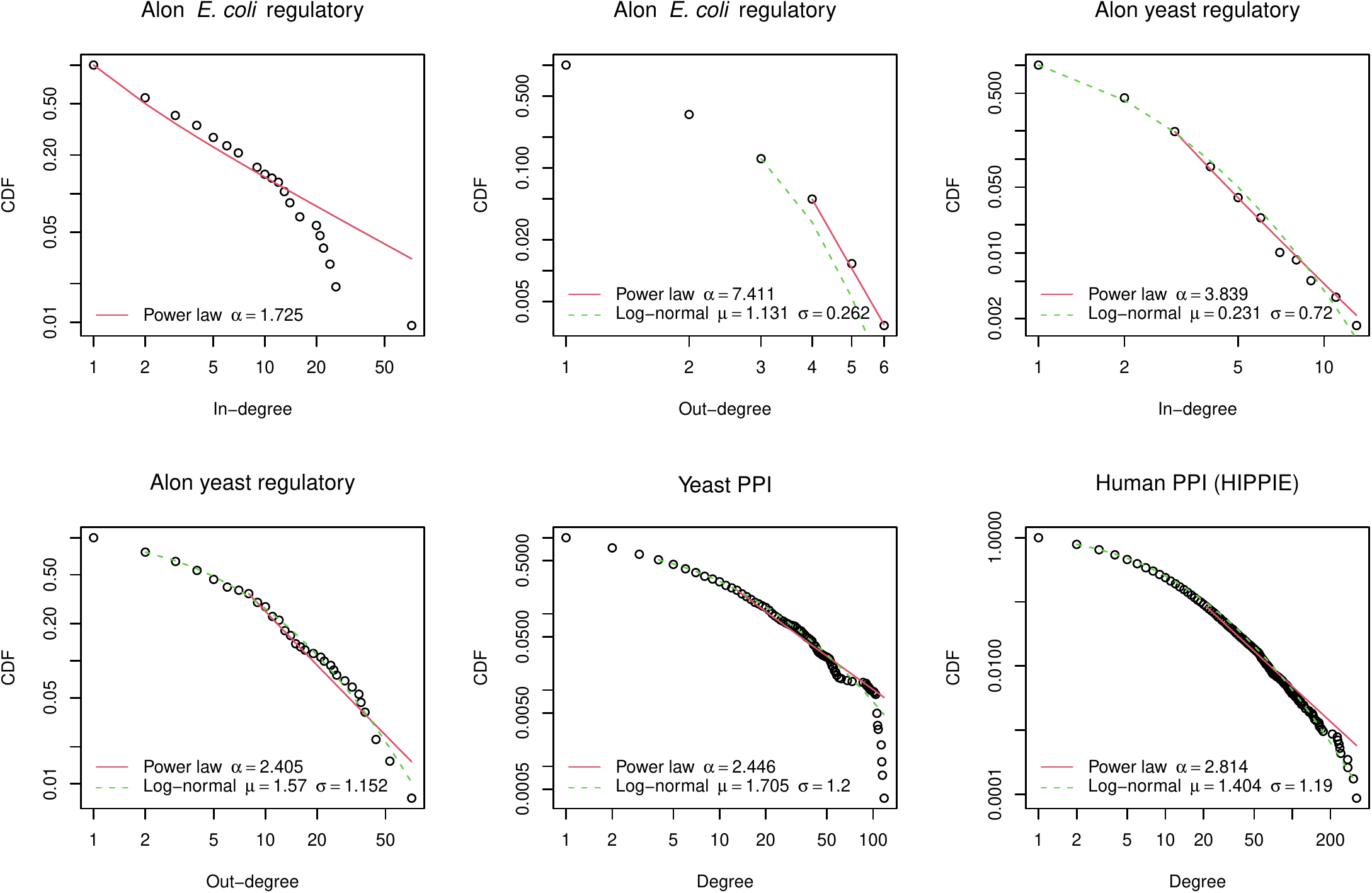}
  \caption{Degree distributions of the networks.
    Power law and
    log-normal distributions fitted to the CDF for
    degree distributions of the networks (in- and out-degree for directed networks, degree for undirected networks).
    All distributions apart from the \textit{E. coli} in-degree distribution
    (for which a log-normal distribution could not be fitted), 
    and the Human PPI (HIPPIE) degree distribution (which is not consistent
    with a power law distribution, $p < 0.01$),
    are consistent with both power law and log-normal distributions
  \label{fig:degree_distributions}}
\end{figure}

Summary statistics of the networks are in
Table~\ref{tab:bionetworks_graph_stats} and the degree distributions of
the networks are shown in Fig.~\ref{fig:degree_distributions}.
In this figure, $\alpha$ is the exponent in the discrete power
law distribution $\Pr(X=x) = Cx^{-\alpha}$ (where $C$ is a
normalization constant), and $\mu$ and $\sigma$ are the parameters
(respectively, mean and standard deviation of $\log(x)$) of the
discrete log-normal distribution.  Power law and log-normal
distributions were fitted using the methods of
\citet{clauset09} implemented in the poweRlaw package
\citep{gillespie15}.

\begin{table}
  \caption{Summary statistics for the biological networks
    \label{tab:bionetworks_graph_stats}}
    {\begin{tabular*}{\textwidth}{@{\extracolsep{\fill}}llrrrr@{}}
      \toprule
      Network & Directed & Nodes  &   Edges     &    Density & Clustering  \\
      &          &                &             &            &  coefficient \\
      \midrule
      Yeast PPI & No & 2617  & 11855 & 0.00346 & 0.46862 \\
      Human PPI (HIPPIE) & No & 11517 & 47184 & 0.00071 & 0.03765 \\
      Alon \textit{E.~coli} regulatory & Yes  & 423 & 519 & 0.00291 & 0.02382 \\
      Alon yeast regulatory & Yes & 688 & 1079 & 0.00228 & 0.01625 \\
      \bottomrule
    \end{tabular*}}

    \parbox{\textwidth}{``Clustering coefficient'' is the global clustering
    coefficient (transitivity)}
\end{table}

\subsection*{ERGM configurations}

\begin{figure}
  \centering
  \includegraphics[width=0.9\textwidth]{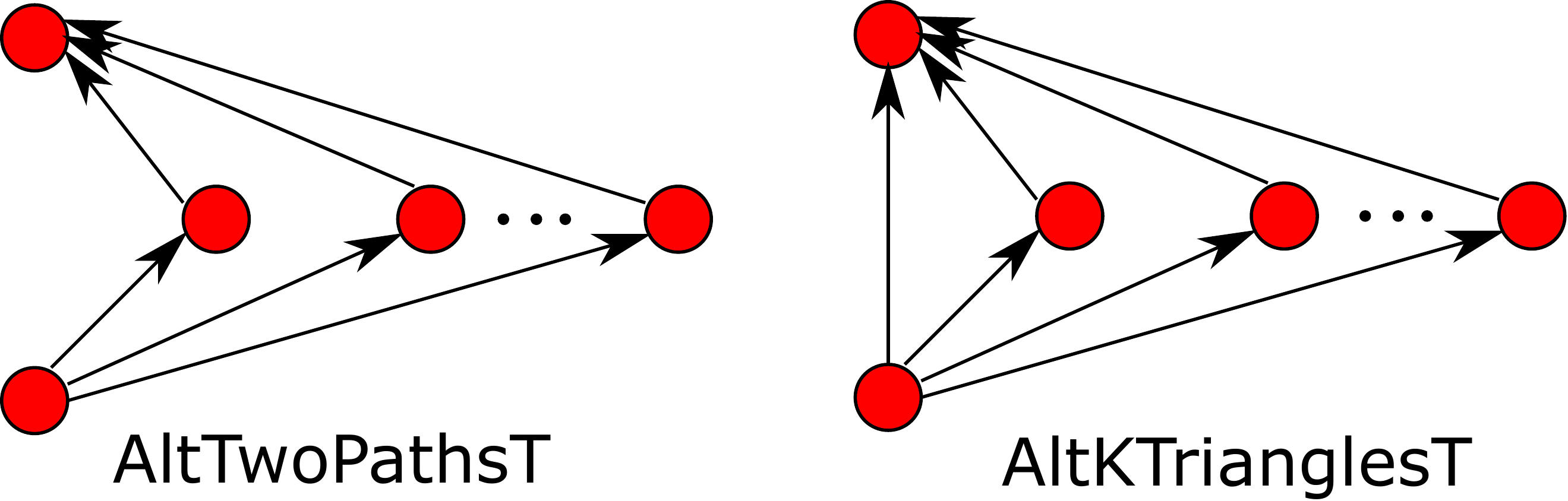}
  \caption{Alternating two-paths and alternating
      transitive triangles ERGM configurations for directed
      networks. Unlike motifs, ERGM configurations are not induced
    subgraphs, so it is normal (and often required) for one to be a
    subgraph of another. So AltTwoPathsT and AltKTrianglesT are
    frequently included in a model together, with AltKTrianglesT
    consisting of the AltTwoPathsT configuration ``closed'' by the
      addition of an arc
  \label{fig:alternating_configurations}}
\end{figure}

The ERGM parameters used in the models for undirected networks are
shown in Table~\ref{tab:undirected_terms}, and those for directed
networks in Table~\ref{tab:directed_terms}. Detailed descriptions of
these parameters and their corresponding statistics can be found in
\citet{robins07intro,lusher13,snijders06,robins07,robins09,stivala20},
but two of the important ones used in this work are shown in
Fig.~\ref{fig:alternating_configurations}.

\begin{table}[!t]
  \caption{Parameters for undirected networks
    \label{tab:undirected_terms}}
  {\begin{tabular*}{\textwidth}{@{\extracolsep{\fill}}lp{0.7\linewidth}@{}}
      \toprule
      Effect & Description \\
      \midrule
      Edge &  Baseline density. \\
      A2P & Alternating $k$-two-paths. Used as a ``control'' for alternating $k$-triangles. \\
      AS & Alternating $k$-stars. A positive parameter value indicates centralization based on high-degree nodes.\\
      AT &  Alternating $k$-triangles. A positive parameter value indicates network closure (triangles). \\
      Match $c$ & Categorical matching on categorical attribute $c$. A positive parameter value indicates an edge preferentially forming between nodes with the same value of the categorical attribute (known as ``homophily'' in social network research). \\
      \bottomrule
  \end{tabular*}}{}
\end{table}

\begin{table}
  \caption{Parameters for directed networks
    \label{tab:directed_terms}}
  {\begin{tabular*}{\textwidth}{@{\extracolsep{\fill}}lp{0.7\linewidth}@{}}
      \toprule
      Effect & Description \\
      \midrule
      Arc & Baseline density. \\
      Sink & A positive parameter value indicates a tendency for nodes with incoming but no outgoing arcs. \\
      Source & A positive parameter value indicates a tendency for nodes with outgoing but no incoming arcs.  \\
      Reciprocity & A positive parameter value indicates a tendency for arcs to be reciprocated (a cycle of length 2). \\
      AltInStars &  Alternating $k$-in-stars. A positive parameter value indicates centralization based on high in-degree nodes. \\
      AltOutStars & Alternating $k$-out-stars. A positive parameter value indicates centralization based on high out-degree nodes. \\
      AltTwoPathsT & Multiple 2-paths. A positive parameter value indicates a tendency for directed paths of length 2. Used as a ``control'' for AltKTrianglesT, the parameter for triangles formed by closing these 2-paths. \\
      AltKTrianglesT & Path closure or transitive closure. A positive parameter value indicates a tendency for open directed two-paths to be closed transitively. This is an alternating statistic version of the ``feed-forward loop'' motif. \\
      AltKTrianglesC & Cyclic closure. A positive parameter value indicates a tendency for directed cycles of length 3 in the network, representing non-hierarchical network closure. An alternating statistic version of the ``three-node feedback loop'' motif.  \\
      Sender $a$ & Sender on binary attribute $a$. A positive parameter value indicates that nodes with the attribute are more likely to have an incident arc directed from them. \\
      Receiver $a$ & Receiver on binary attribute $a$. A positive parameter value indicates that nodes with the attribute are more likely to have an incident arc directed to them. \\
      Interaction $a$ & Interaction on binary attribute $a$. A positive parameter value indicates that two nodes which both have the attribute are more likely to have an arc directly connecting them. \\
      Matching $c$ & Matching on categorical attribute $c$. A positive parameter value indicates that two nodes which have the same value of the attribute are more likely to have an arc directly connecting them.  \\
      Loop & Self-edge. A positive parameter value indicates a tendency for self-edges (loops). \\ 
    \bottomrule
  \end{tabular*}}{}
\end{table}

The ``alternating'' statistics \citep{snijders06,robins07,lusher13}
such as alternating $k$-stars involve sums of counts of configurations
with alternating signs and a decay factor $\lambda$, and, except where
otherwise specified, we set $\lambda = 2$ in accordance with common
ERGM modeling practice.

\subsection*{ERGM parameter estimation}

ERGM parameters for undirected networks were estimated using the EE
algorithm \citep{byshkin18} with the IFD sampler \citep{byshkin16}
implemented for undirected networks in the Estimnet software as
described in \citet{byshkin18}, with 20 estimations (run in parallel). ERGM
parameters for directed networks were estimated using the simplified
EE algorithm \citep{byshkin18,borisenko19} with IFD sampler
implemented for directed networks in the EstimNetDirected software
\citep{stivala20}, with 64 estimations (run in parallel).

The Alon \textit{E.~coli} network does not contain any reciprocated
arcs (directed loops of length two), and so estimation is made conditional
on this by preventing the creation of reciprocated arcs in the 
MCMC procedure.

\subsection*{Convergence and goodness-of-fit tests}

Convergence was tested as described in \citet{byshkin18,stivala20}, by requiring
the absolute value of each parameter's t-ratio to be no greater than 0.3,
and by visual inspection of the parameter and statistic trace plots.
For the directed networks estimated with EstimNetDirected, an additional
heuristic convergence test was used, as described in \citet{stivala20}.
Observed graph statistics were plotted on the same plots as the distributions
of those statistics in the networks simulated in the EE algorithm MCMC
process, to check that they do not diverge. The statistics used are the
same as those of the actual goodness-of-fit test described below, but
note that this test is only for estimation
convergence, not goodness-of-fit \citep{stivala20}.

For the directed networks estimated with EstimNetDirected, a
simulation-based goodness of fit procedure was used, similar to
that used in statnet \citep{hunter2008ergm}. A set of networks was
simulated from the estimated model (using the SimulateERGM program in the EstimNetDirected software), and the distribution of certain
graph statistics compared with those of the observed network by
plotting the observed network values on the same plots as the
distribution of simulated values. The statistics used were the in- and
out-degree distributions, reciprocity, giant component size, mean
local and global clustering coefficients, triad census, geodesic distance
(shortest path length) distribution, and edge-wise and dyad-wise shared
partners distributions.

\section*{Results and discussion}

Table~\ref{tab:yeast_ee_models} shows the basic structural model for
the yeast PPI network (Model 1), a model with the alternating
$k$-two-paths (A2P) parameter added (Model 2), as well as a model
(Model 3) incorporating a parameter for the propensity of interactions
to occur between proteins in the same functional category
(class). Model 1 reproduces a model of this network in a previous work
\citep[Table~S3]{byshkin18}; Models 2 and 3 are new.

\begin{table}
  \caption{Parameter estimates with 95\% confidence interval
    for the yeast PPI network, from the EE algorithm
    \label{tab:yeast_ee_models}}
    {\begin{tabular*}{\textwidth}{@{\extracolsep{\fill}}lrrr@{}}
      \toprule
      Effect  & Model 1 & Model 2 & Model 3\\
      \midrule
      Edge  & $\heavy{\underset{(-7.806, -7.709)}{-7.758}}$ & $\heavy{\underset{(-10.685, -10.650)}{-10.667}}$ & $\heavy{\underset{(-9.302, -9.262)}{-9.282}}$\\
      AS  & ${\underset{(-0.103, 0.007)}{-0.048}}$ & $\heavy{\underset{(1.013, 1.140)}{1.077}}$ & $\heavy{\underset{(0.550, 0.659)}{0.604}}$\\
      A2P  & --- & $\heavy{\underset{(-0.090, -0.084)}{-0.087}}$ & $\heavy{\underset{(-0.062, -0.056)}{-0.059}}$\\
      AT  & $\heavy{\underset{(1.807, 1.907)}{1.857}}$ & $\heavy{\underset{(2.474, 2.548)}{2.511}}$ & $\heavy{\underset{(2.396, 2.467)}{2.432}}$\\
      Match class  & --- & --- & $\heavy{\underset{(0.315, 0.402)}{0.358}}$\\
      \bottomrule
    \end{tabular*}}

    \parbox{\textwidth}{Parameter estimates that are statistically
        significant are shown in bold}
\end{table}

Each of these model estimations took approximately 7 minutes total
elapsed time on cluster nodes with Intel Xeon E5-2650 v3 2.30GHz
processors using 20 parallel tasks.

We expect that proteins of the same functional
category should preferentially interact with each other
\citep{vonmering02}, and this is confirmed by the significant positive
parameter estimated for the ``Match class'' effect. The alternating
$k$-triangle (AT) parameter is positive and significant in all models,
showing an over-representation of triangles (which we might
expect given the very high value of the clustering coefficient for this
network, Table~\ref{tab:bionetworks_graph_stats}), even in models also
including parameters for two-paths and preferential interaction of
proteins in the same class.

Table~\ref{tab:hippie_ee_models} shows a basic structural model for
the human PPI high confidence network (Model 1), and a model with a
term to control for subellular location by categorical matching on the
cellular component GO term (Model 2).

\begin{table}
  \caption{Parameter estimates with 95\% confidence interval
    for the human PPI (HIPPIE high confidence) network, from the EE algorithm
    \label{tab:hippie_ee_models}}
    {\begin{tabular*}{\textwidth}{@{\extracolsep{\fill}}lrrr@{}}
     \toprule
     Effect  & Model 1 & Model 2\\
     \midrule
     Edge  & $\heavy{\underset{(-12.550, -12.450)}{-12.500}}$ & $\heavy{\underset{(-12.498, -12.402)}{-12.450}}$\\
     AS  & $\heavy{\underset{(1.222, 1.258)}{1.240}}$ & $\heavy{\underset{(1.208, 1.245)}{1.226}}$\\
     A2P  & $\heavy{\underset{(-0.001, -0.001)}{-0.001}}$ & $\heavy{\underset{(-0.001, -0.001)}{-0.001}}$\\
     AT  & $\heavy{\underset{(1.692, 1.711)}{1.701}}$ & $\heavy{\underset{(1.680, 1.699)}{1.690}}$\\
     Match cellular component  & --- & $\heavy{\underset{(0.431, 0.498)}{0.465}}$\\
     \bottomrule
    \end{tabular*}}

    \parbox{\textwidth}{Parameter estimates that are statistically
        significant are shown in bold}
\end{table}

Estimation of Model 1 took approximately 64 minutes elapsed time, and
Model 2 approximately 73 minutes, on cluster nodes with Intel Xeon
E5-2650 v3 2.30GHz processors using 20 parallel tasks.

As discussed in the Introduction, we expect that interactions would be
over-represented between proteins that share a subcellular location,
and this is confirmed by a statistically significant positive
parameter estimate for categorical matching on cellular component
(Model 2 in Table~\ref{tab:hippie_ee_models}).  The alternating
$k$-triangle (AT) parameter is positive and statistically significant
in both models. This indicates an over-representation of triangles,
even when controlling for subcellular location (Model 2).

We estimated four different models of the Alon \textit{E.~coli}
regulatory network (Table~\ref{tab:alon_ecoli_ergm}). In Models 1 and 2, following
\citet{hummel12}, we modeled self-regulation by using a nodal
covariate ``self'' which is true exactly when the node had a self-edge (loop)
in the original network. These ERGM models are new, in that previous
work with ERGMs on these networks either treated them as undirected
\citep{saul07,hummel12}, thereby ignoring the inherently directed
nature of such a regulatory network; or, in the case where the
network was left as directed, included only Arc and alternating $k$-in-stars terms,
as the estimation methods used at the time could not find converged
models when other terms, such as triangles, were included
\citep{begum14}.

\begin{table}
   \caption{Parameter estimates with 95\% confidence interval for the
     Alon \textit{E. coli} regulatory network
  \label{tab:alon_ecoli_ergm}}
   {\begin{tabular*}{\textwidth}{@{\extracolsep{\fill}}lrrrr@{}}
\toprule
Effect  & Model 1 & Model 2 & Model 3 & Model 4\\
\midrule
Arc  & $\heavy{\underset{(-8.368, -8.047)}{-8.208}}$ & $\heavy{\underset{(-7.830, -7.509)}{-7.670}}$ & $\heavy{\underset{(-5.535, -5.149)}{-5.342}}$ & $\heavy{\underset{(-8.664, -8.337)}{-8.500}}$\\
Sink  & $\light{\underset{(-0.240, 6.830)}{3.295}}$ & $\heavy{\underset{(0.306, 5.448)}{2.877}}$ & $\light{\underset{(-5.584, 4.330)}{-0.627}}$ & $\light{\underset{(-1.512, 9.662)}{4.075}}$\\
Source  & $\light{\underset{(-1.770, 4.247)}{1.238}}$ & $\light{\underset{(-1.063, 4.156)}{1.546}}$ & $\light{\underset{(-3.084, 5.195)}{1.056}}$ & $\light{\underset{(-1.683, 6.551)}{2.434}}$\\
AltInStars  & $\heavy{\underset{(0.943, 4.232)}{2.587}}$ & $\heavy{\underset{(0.555, 4.044)}{2.299}}$ & $\light{\underset{(-0.163, 3.392)}{1.614}}$ & $\heavy{\underset{(0.630, 4.798)}{2.714}}$\\
AltOutStars  & $\light{\underset{(-2.363, 0.362)}{-1.001}}$ & $\light{\underset{(-1.885, 0.189)}{-0.848}}$ & $\light{\underset{(-3.987, 1.717)}{-1.135}}$ & $\light{\underset{(-2.300, 1.266)}{-0.517}}$\\
AltTwoPathsT  & $\light{\underset{(-0.638, 0.297)}{-0.170}}$ & $\light{\underset{(-0.603, 0.272)}{-0.165}}$ & $\light{\underset{(-1.376, 0.192)}{-0.592}}$ & $\light{\underset{(-0.767, 0.267)}{-0.250}}$\\
AltKTrianglesT  & $\heavy{\underset{(0.798, 4.972)}{2.885}}$ & $\heavy{\underset{(1.025, 4.636)}{2.830}}$ & $\heavy{\underset{(0.722, 5.555)}{3.139}}$ & $\heavy{\underset{(0.548, 5.604)}{3.076}}$\\
Matching self  & --- & $\light{\underset{(-1.181, 0.280)}{-0.451}}$ & --- & ---\\
Loop  & --- & --- & --- & $\heavy{\underset{(1.949, 14.256)}{8.103}}$\\
\bottomrule
   \end{tabular*}}

    \parbox{\textwidth}{Parameter estimates that are statistically significant are shown in bold. In Models 3 and 4, self-edges
    (loops) are retained in the network and allowed in the model}
\end{table}

Each of these model estimations took approximately three minutes total
elapsed time on cluster nodes with Intel Xeon E5-2650 v3 2.30GHz
processors using 64 parallel tasks.

\begin{figure}
  \centering
  \includegraphics[width=\textwidth]{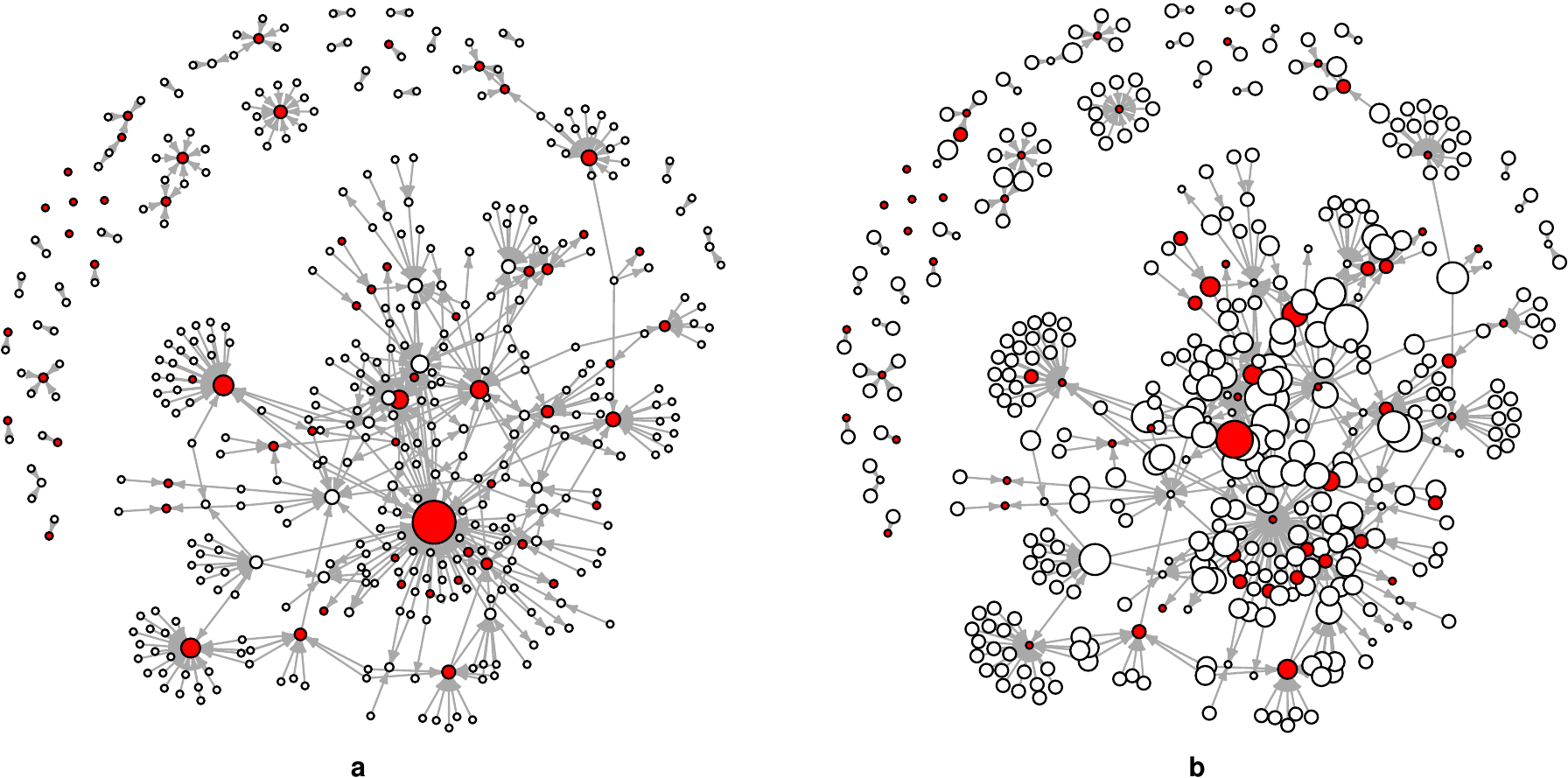}
  \caption{Alon \textit{E.~coli} regulatory network.
    (\textbf{a}) Node
    size is proportional to in-degree. (\textbf{b}) Node size is
    proportional to out-degree. Self-regulating operons are depicted
    as filled (red) circles.  In (a) there appears to be a small set
    of high in-degree nodes and a much larger set of smaller in-degree
    nodes, while in (b) the out-degree of the nodes appears to be much
    more evenly distributed. The hypothesis we might make from (a),
    that there is centralization on in-degree, is confirmed by the
    ERGM results. This same model finds no support for the
    hypothesis we might make from (b), that there is a tendency against
    centralization on out-degree       
  \label{fig:alon_ecoli_network}}
\end{figure}

In these models, the Sink and Source parameters are used to control,
respectively, for the presence of genes that do not regulate any genes
(have out-degree zero) and genes that are not regulated by any gene
(have in-degree zero). The alternating $k$-in-stars (AltInStars)
parameter is positive and significant in all models except Model 3, indicating significant
skewness of the in-degree distribution, that is, the presence of ``hubs'' with
higher in-degree than other nodes. There is no significant
effect for (or against) such skewness of the out-degree distribution (see
Fig.~\ref{fig:degree_distributions} and
Fig.~\ref{fig:alon_ecoli_network}).

The only other parameter that is consistently significant (and
positive) is path closure (AltKTrianglesT), which we can
interpret as a significant tendency for the ``feed-forward loop'' to be
over-represented, consistent with the results in \citet{milo02}.

\begin{figure}
  \centering
  \includegraphics[width=0.8\textwidth]{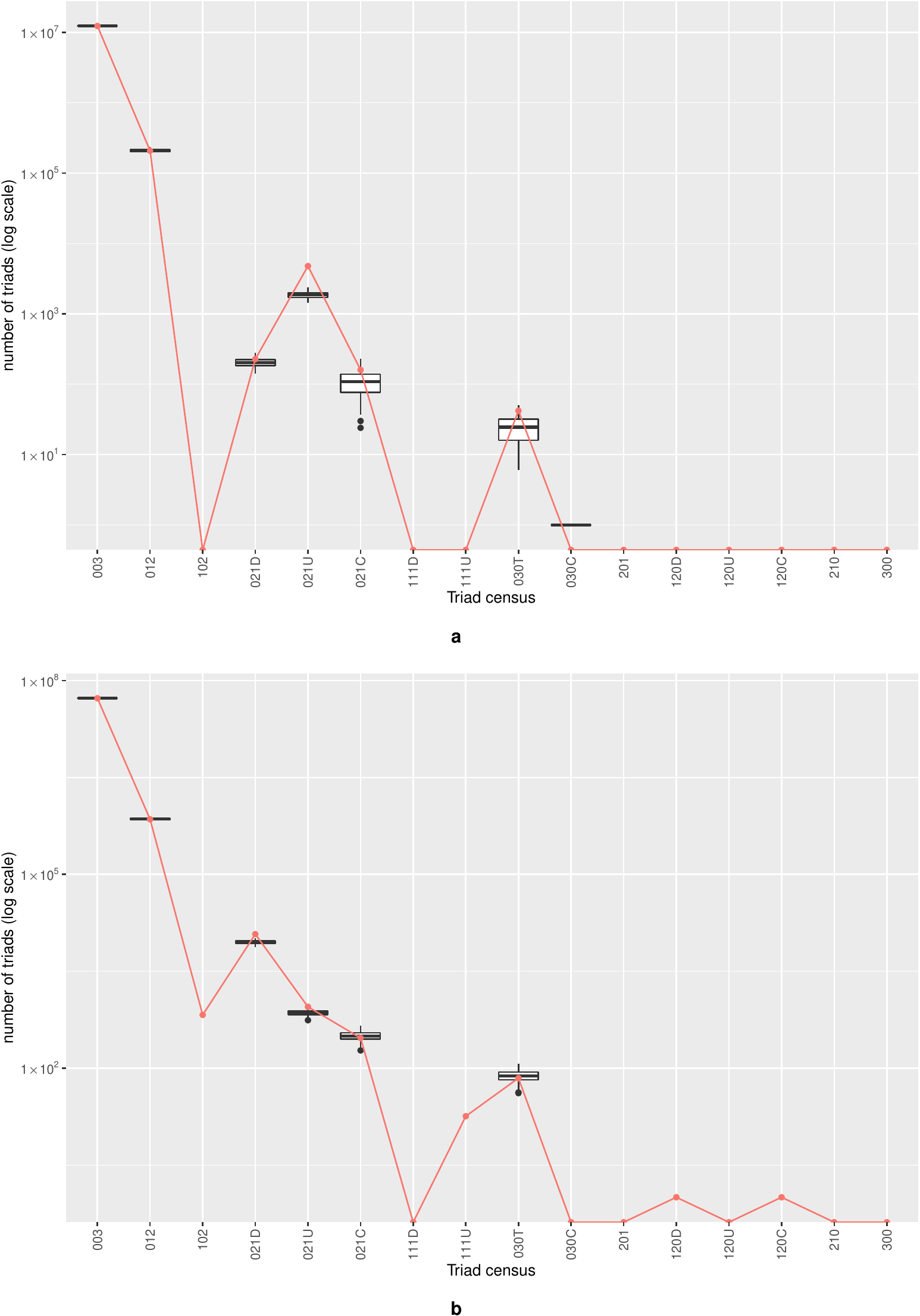}
  \caption{Goodness-of-fit plots for the triad census of (\textbf{a}) the Alon
    \textit{E. coli} regulatory network, Model 1
    (Table~\ref{tab:alon_ecoli_ergm}), and (\textbf{b})
    the Alon yeast regulatory network, Model 1
    (Table~\ref{tab:alon_yeast_lambda_ergm}).
    The observed triad counts are
    plotted in red with the triad counts of 100 simulated networks
    plotted as black boxplots.
    Because the triad counts ($y$-axis)
    are on a log scale, values of zero are omitted (observed zero
    counts shown as a red point on the bottom of the graph).
    In (\textbf{a}), for
    triad census class 030C (cyclic triad), the ``box plot'' consisting of a single median line for
    the simulated count represents a single (out of 100
    simulations) occurrence of a nonzero count (of 1) for 030C
    \label{fig:alon_ecoli_yeast_gof_triadcensus}}
\end{figure}

A goodness-of-fit plot for Model 1 (Table~\ref{tab:alon_ecoli_ergm})
is shown in \SI~(Fig.~S1(a)), showing a good fit for the model.  A
goodness-of-fit plot for the triad census
(Fig.~\ref{fig:alon_ecoli_yeast_gof_triadcensus}(a)) shows that the
model reproduces the triad census well, and specifically triad 030T,
the transitive triad (three node feed-forward loop), giving additional
confidence that the positive and statistically significant
AltKTrianglesT parameter is evidence for over-representation of this
motif, given the other parameters in the model.

Note that this \textit{E.~coli} regulatory network does not contain
any instances of the three-cycle, or ``three-node feedback loop''
\citep{milo02}. Indeed the Alon \textit{E.~coli} network does not
contain any loops greater than size one \citep{shen02}, and so the
cyclic closure parameter (AltKTrianglesC) is not included in the
models.

In Models 3 and 4 (Table~\ref{tab:alon_ecoli_ergm}), unlike the other
models, self-edges (loops) are retained in the network, and self-edges
are allowed in the modeling process, allowing the formation of loops
to be modeled jointly with the other structural features in the model.
\footnote{Modeling self-edges in this way was suggested by
an anonynous reviewer, on the grounds that a node with a self-edge
is a (very simple) motif.}
In Model 4, the new parameter ``Loop'' is introduced, for which the
corresponding statistic is the count of self-edges in the
network. This parameter is statistically significant and positive,
indicating that self-edges are over-represented, given the other
effects included in the model.  Goodness-of-fit plots for Models 3 and
4 (Table~\ref{tab:alon_ecoli_ergm}) are shown in \SI, Fig.~S4, showing
that when the Loop parameter is not included in the model (Model 3 in
Table~\ref{tab:alon_ecoli_ergm}), there is a poor fit for the number
of loops (\SI, Fig.~S4(a)). However, when the Loop parameter is
included (Model 4 in Table~\ref{tab:alon_ecoli_ergm}), there is a
good fit for the number of loops (\SI, Fig.~S4(b)).

We found that it is also possible to estimate similar models of this
relatively small network using the most recent version of the statnet
ergm package \citep{ergm,krivitsky21}, with the ``stepping'' algorithm
\citep{hummel12}. These models are shown in \SI~(Table~S1), and
the goodness-of-fit plots in \SI~(Fig.~S6 and Fig.~S7). The
results are consistent with those in
Table~\ref{tab:alon_ecoli_ergm}. Specifically, there is a significant
positive estimate for geometrically weighted edge-wise shared partners
(GWESP, equivalent to AltKTrianglesT), and a significant negative
estimate for geometrically weighted in-degree, indicating
centralization in the in-degree distribution.\footnote{Note that a
\emph{negative} estimate of the geometrically weighted degree
parameter in statnet has the same interpretation as a \emph{positive}
estimate for the alternating $k$-stars parameter as used here. This
frequently leads to confusion \citep{levy16,levy16poster}.}  The
statnet model finds a significant tendency against centralization on
out-degree, while the models in Table~\ref{tab:alon_ecoli_ergm} did
not have a significant estimate for the corresponding parameter
(AltOutStars). Similarly the statnet model (Model 2 in \SI, Table~S1)
finds a significant negative parameter estimate for Matching on the
``self-regulating'' attribute, while no significant effect is found in
Model 2 in Table~\ref{tab:alon_ecoli_ergm}.  The statnet ergm
package does not allow for the modeling of self-edges, however
\citep{hummel12}.

Table~\ref{tab:alon_yeast_lambda_ergm} shows ERGM parameter estimates for the
Alon yeast regulatory network.  Each of these model estimations took
approximately three minutes total elapsed time on cluster nodes with
Intel Xeon E5-2650 v3 2.30GHz processors using 64 parallel
tasks. These ERGM models are also new; previously published ERGMs for
similar networks having treated them as undirected \citep{saul07}.

\begin{table}
  \caption{Parameter estimates with 95\% confidence intervals for the
    Alon yeast regulatory network
    \label{tab:alon_yeast_lambda_ergm}}
  {\begin{tabular*}{\textwidth}{@{\extracolsep{\fill}}lrrr@{}}
      \toprule
      Effect  & Model 1 & Model 2 & Model 3\\
      \midrule
      Arc  & $\heavy{\underset{(-7.665, -7.313)}{-7.489}}$ & $\heavy{\underset{(-7.662, -7.318)}{-7.490}}$ & $\heavy{\underset{(-7.667, -7.313)}{-7.490}}$\\
      Reciprocity  & --- & --- & $\light{\underset{(-15.535, 3.307)}{-6.114}}$\\
      AltInStars   & $\light{\underset{(-1.504, 0.577)}{-0.463}}$ & $\light{\underset{(-1.451, 0.581)}{-0.435}}$ & $\light{\underset{(-1.472, 0.568)}{-0.452}}$\\
      AltOutStars ($\lambda = 4.5$)  & $\heavy{\underset{(0.756, 1.261)}{1.008}}$ & $\heavy{\underset{(0.757, 1.256)}{1.006}}$ & $\heavy{\underset{(0.752, 1.264)}{1.008}}$\\
      AltTwoPathsT ($\lambda = 3$)  & $\light{\underset{(-0.739, 0.076)}{-0.332}}$ & $\light{\underset{(-0.670, 0.074)}{-0.298}}$ & $\light{\underset{(-0.684, 0.054)}{-0.315}}$\\
      AltKTrianglesT ($\lambda = 3$)  & $\heavy{\underset{(0.484, 4.111)}{2.297}}$ & $\heavy{\underset{(0.043, 3.640)}{1.842}}$ & $\heavy{\underset{(0.479, 3.763)}{2.121}}$\\
      \bottomrule
  \end{tabular*}}

  \parbox{\textwidth}{Parameter estimates that are statistically
    significant are shown in bold.  In Model 1 only, estimation is
    conditional on no reciprocated arcs, even though there is a single
    reciprocated arc in the data. Model 3 is included for
    illustration, even though it shows poor convergence with respect
    to the Reciprocity parameter (t-ratio magnitude is greater than
    0.3)}
\end{table}

In Model 1 (Table~\ref{tab:alon_yeast_lambda_ergm}), estimation is
conditional on no reciprocated arcs, just as was done for the
\textit{E.~coli} regulatory network. However in this yeast regulatory
network, there is actually a single reciprocated arc (two-cycle) in
the data, and hence the fit of the model on statistics involving
reciprocated arcs is poor. This is apparent, for example, in the poor
fit for triad census class 102 (triad with only a mutual arc) in
Fig.~\ref{fig:alon_ecoli_yeast_gof_triadcensus}(b), or for the
reciprocity statistic in the goodness-of-fit plot
(\SI, Fig.~S1(b)). The fit for other statistics, and in
particular the degree and shared partner distributions, is acceptable
(with the exception of poor fit on the giant component size).
Importantly, the fit on the triad census class 030T (transitive triad)
is good (Fig.~\ref{fig:alon_ecoli_yeast_gof_triadcensus}(b)).

In order to better model reciprocity, a model (Model 2 in
Table~\ref{tab:alon_yeast_lambda_ergm}) was estimated without being
conditional on there being no reciprocated arcs, but without a
reciprocity term in the model.  This model also has adequate
goodness-of-fit, but this time including good fit on the reciprocity
statistic (\SI, Fig.~S2(a)). It does,
however, for some triads involving reciprocated arcs (120U for
example), generate significantly more such triads than are observed in
the data
(\SI, Fig.~S2(b)). Therefore, a
third model (Model 3 in Table~\ref{tab:alon_yeast_lambda_ergm}) was
estimated, including the Reciprocity parameter. However, probably due
to the fact that the data contains only a single reciprocated arc,
this model has a very large estimated standard error for the
Reciprocity parameter. Further, it exhibits poor convergence with
respect to the Reciprocity statistic, with a t-ratio greater than the
maximum value of 0.3 we consider acceptable, since the data contains
exactly one reciprocated arc, yet the model most frequently generates
networks with none.

Model 1 and Model 2, therefore, are preferable. Nevertheless, in all three
models, the sign and significance of estimated parameters (except Reciprocity)
are the same. There is a positive and significant parameter for
alternating $k$-out-stars (AltOutStars), indicating the presence
of ``hubs'' with higher out-degree than other nodes. This is as we might
expect from Fig.~\ref{fig:degree_distributions} and previous
research \citep{monteiro20,guelzim02,balaji06,ouma18}, and contrasts
with the \textit{E.~coli} regulatory network, which has in-degree hubs
but not out-degree hubs.

Also in all three models, there is a positive and significant
parameter estimate for transitive closure (AltKTrianglesT). Given this
estimate, and the good fit for the transitive closure motif 030T
(Fig.~\ref{fig:alon_ecoli_yeast_gof_triadcensus}(b)) we can again
interpret this as a significant over-representation of this motif
(``feed-forward loop''), consistent with the results of
\citet{milo02}.

In all three models in Table~\ref{tab:alon_yeast_lambda_ergm}, the decay
parameter $\lambda$ for the ``alternating'' statistics has been set to
a value other than the default $\lambda=2$ for alternating
$k$-out-stars (AltOutStars), multiple two-paths (AltTwoPathsT), and
transitive closure (AltKTrianglesT). This is because models initially
estimated with the default $\lambda=2$ value
(\SI, Table~S2) showed poor goodness-of-fit on the
out-degree distribution (\SI, Fig.~S3(a)) and
triad census class 030T
(\SI, Fig.~S3(b)).  Therefore, new
models were estimated with a higher value of $\lambda$ for the
alternating $k$-out-star parameter to assist with modeling the highly
skewed out-degree distribution \citep{koskinen13}, and also a higher
value of $\lambda$ for AltTwoPathsT and AltKTrianglesT (the same
value of $\lambda$ for both) to aid model convergence and fit for transitivity
\citep{snijders06}.

As with the \textit{E.~coli} network, we also estimated a model of the
yeast regulatory network, in which self-edges are retained, and
allowing self-edges (loops) in the model. This network (even leaving
aside the presence of self-edges) is, however, not identical to the
network used for the models shown in
Table~\ref{tab:alon_yeast_lambda_ergm}, having two additional nodes.
Its graph summary statistics, are, however the same (to the precision shown) as those of the
version shown in Table~\ref{tab:bionetworks_graph_stats}, other than
it having 690 rather than 688 nodes. Since the network modeled is a
slightly different network than that used for the models shown in
Table~\ref{tab:alon_yeast_lambda_ergm}, these models are presented separately,
in \SI, Table~S3. The results are consistent with those in
Table~\ref{tab:alon_yeast_lambda_ergm}, with statistically significant
positive parameter estimates for AltOutStars and AltKTrianglesT. The
estimate for the Loop parameter is not statistically significant,
however.  Goodness-of-fit plots for the models in \SI, Table~S3 are
shown in \SI, Fig.~S5. These figures show that the model
which allows self-edges, but does not include the Loop parameter
(Model~1 in \SI, Table~S3) does not fit the number of loops well,
while the model that includes the Loop parameter
(Model~2 in \SI, Table~S3) does fit the number of loops well.

The cyclic triangle structure has been suggested as an ``anti-motif''
(\ie occurs less frequently than expected), but in some cases its
apparent under-representation has been shown to be an expected
consequence of other topological properties of biological networks
\citep{konagurthu08a}. This closed-loop structure, also known as a
``multicomponent loop'', can provide feedback control and potentially
produce systems that can switch between two states
\citep{lee02,ferrel02}.  In the examples used here, there were so few
(or no) occurrences of this motif, that models including the
corresponding parameter (in the form of the AltKTrianglesC parameter)
would not converge. Yet the networks simulated from these models also
contain no (or very few) occurrences of this candidate
anti-motif. This is consistent with the lack of cyclic triangles not
being due to cyclic triangles being an anti-motif as such, but rather
as a consequence of the other topological features of the network, and
specifically in these examples, the features described by the
parameters included in the models. This is not a new finding, it having
previously been noted that the lack of three-node feedback loops in
the \textit{E.~coli} regulatory network \citep{shen02,lee02} is reproduced
in randomized networks \citep{shen02}.

The biological significance of the feed-forward loop (transitive
triangle) is suggested to be that, by providing two
pathways to affect the output, one direct, and one through an
intermediate link, it can act as a logical ``AND'' gate, and filter
out transient activation signals
\citep{shen02,mangan03,alon07,lesk20}. Whether or not this is indeed
the biological function of the feed-forward loop \citep{mazurie05},
this motif is found to be significantly over-represented in the
transcriptional regulatory networks of several organisms
\citep{alon07}, including the yeast and \textit{E.~coli} networks
studied here, and the feed-forward loop has been described as ``highly
favored during the evolution of transcriptional regulatory networks in
yeast'' \citep[p.~801]{lee02}.

More recently, there has been interest in trying to understand the
function of motifs by examining higher levels of
structure. \citet{gorochowski18} examine the clustering of motifs,
including the feed-forward loop, and find that a measure of motif
clustering diversity can predict functionally important nodes in the
\textit{E.~coli} metabolic network.  \citet{lesk20} describes how the
local structure of the yeast regulatory network is reconfigured in
different physiological states.

So far we have only discussed results for three-node motifs, such as
the feed-forward loop. We can test for the over-representation of
other motifs, without including parameters for them in the model, by
using the ERGM as the null model against which to compare the count
of the motif in the observed network. This was the technique used
by \citet{felmlee21}, for example.

\begin{figure}
  \centering
  \includegraphics[width=0.6\textwidth]{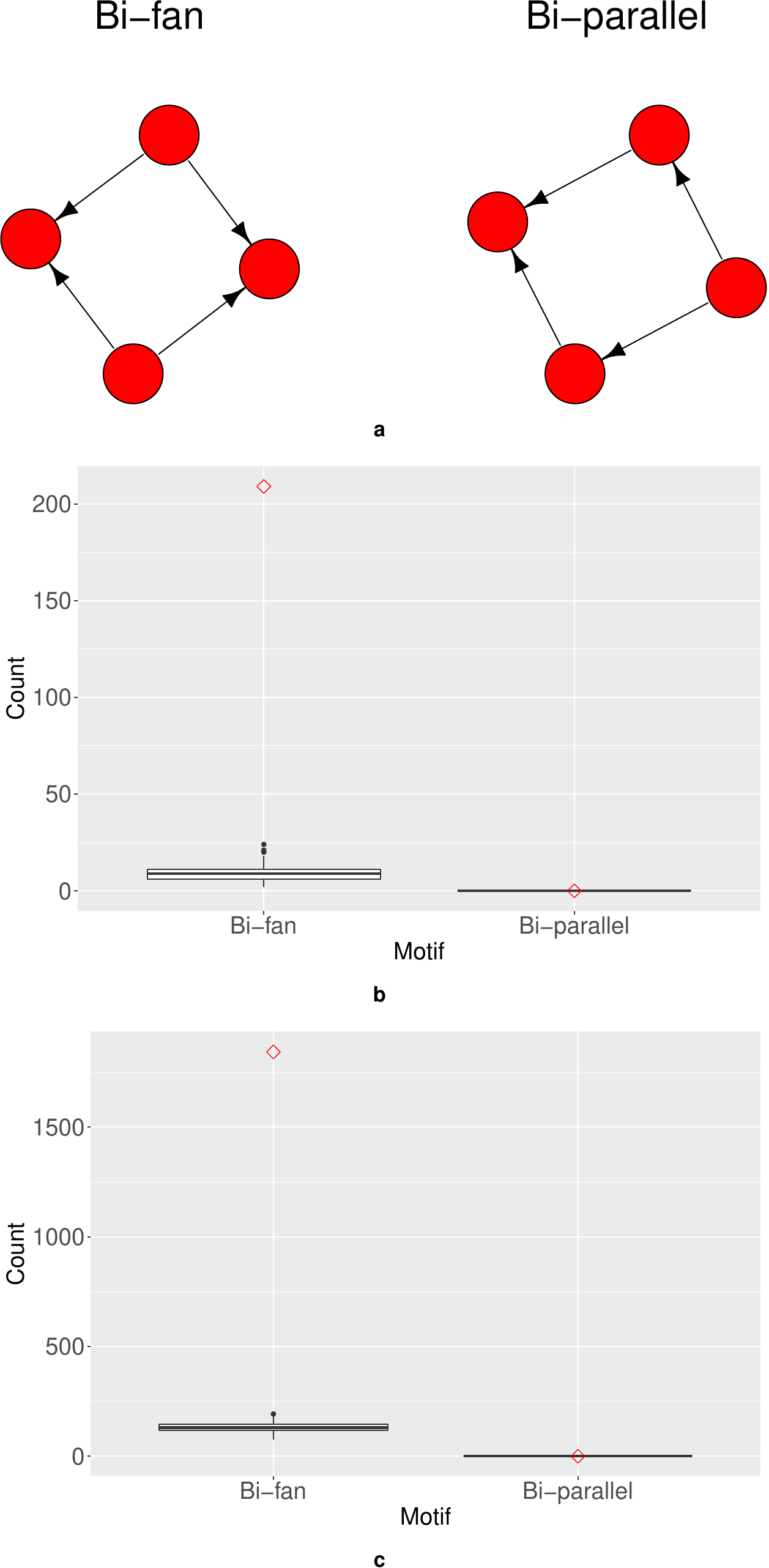}
  \caption{(\textbf{a}) The bi-fan and bi-parallel four-node motifs.
    Goodness-of-fit plots for these motifs
    for (\textbf{b}) the Alon \textit{E. coli} regulatory
    network Model 1 (Table~\ref{tab:alon_ecoli_ergm}), and
    (\textbf{c}) the Alon yeast regulatory network Model 1
    (Table~\ref{tab:alon_yeast_lambda_ergm}).  The observed network
    statistics are plotted as a red diamond, with the statistics of
    100 simulated networks plotted as black boxplots
    \label{fig:tetrad_figures}}
\end{figure}

Fig.~\ref{fig:tetrad_figures} shows the bi-fan and bi-parallel motifs,
as defined by \citet{milo02}, their counts in the \textit{E. coli} and
yeast regulatory networks, and their distribution in ERGM models of
these networks. The motifs were counted with the NetMODE software
\citep{li12}. Note that NetMODE was used only to count the motifs, not
to simulate any networks, which are simulated from the ERGM models as
described in the Methods section.

The bi-parallel motif occurs in neither of the observed networks, and
nor does it occur in any of the networks simulated from the
corresponding ERGMs. The bi-fan motif, however, clearly occurs far
more frequently in both observed networks than it does in the
corresponding simulated networks.  Note that these networks are
simulated from ERGMs that model not just degree distribution, but also
the distribution of two-paths and transitive triangles. Therefore,
this shows that the bi-fan motif appears to be over-represented in the
observed networks, even given the over-representation of transitivity
captured in the models, which also reasonably reproduce the triad
census, geodesic distance distribution, and dyad-wise and edge-wise
shared partner distributions. These results are consistent with the
results of \citet{milo02}, where only degree-preserving randomization
was used.

\section*{Limitations}

Finding a converged ERGM for a network is not always possible in
practice. In particular, models which include Markov dependency
assumption parameters such as triangles, corresponding directly to
three-node motif candidates such as three-node feed-forward-loops
(transitive triangles) and three-cycles, for example, usually do not
converge. For this reason it is normal practice in ERGM modeling to
use geometrically weighted or ``alternating'' configurations to solve
this problem \citep{snijders06,robins07,hunter12}, as we did in this
work. However this means we are not answering precisely the same
question as when we ask directly if a motif is over-represented or
not. This is because ERGM is a model for tie (edge or arc) formation,
not for motif formation: if we consider ERGM as a type of logistic regression,
the outcome variable is the presence or absence of a network tie. The
predictor variables are not independent of each other, but form a
nested hierarchy of configurations: triangles are formed by
``closing'' a two-path with an additional edge, for example. So a
positive estimate of the alternating $k$-triangle parameter does not
directly mean that the transitive triangle (three node feed-forward
loop) motif is over-represented, but rather that there is tendency
(that is, it is more probable than chance given the other parameters
in the model) for three nodes forming a directed two-path to be closed
in a transitive triangle. This makes sense in
the social network origins of the model: it might be assumed to be the
result in the observed network of the tendency of a person's friends
to also be friends with each other, for example. In the context of
biological networks, it might be interpreted as a sign of
evolutionary events, however this interpretation is very much open to
question, as briefly discussed in the Introduction.

Even when the ``alternating'' configurations are used, it can be 
difficult or impossible to find a converged and well-fitting ERGM
for a given network. For example, we were unable to fit an ERGM
with triangular configurations (using either statnet or EstimNetDirected)
to an example of a neural network, the whole-animal chemical connectome (a directed network with 579 nodes and 5~246 arcs) of the male \textit{C.~elegans} worm
\citep{cook19}.

Hence in order to directly test motif significance, without having
to fit a parameterized model such as ERGM, new methods, such as the
``anchored motif'' proposed by \citet{fodor20} are
still required.

In some of the models presented here, we used values other than the
usual default value $\lambda=2$ for the decay parameter $\lambda$ of
the ``alternating'' statistics. We had to manually estimate
appropriate values of $\lambda$ based on trial and error, guided by
knowledge of the observed network, convergence and goodness-of-fit of
the models (or lack thereof), and the definitions of the relevant
statistics \citep{snijders06,koskinen13}. It is possible to instead
estimate $\lambda$ (or an equivalent parameter) directly from the
data, as part of the model, using a ``curved ERGM''
\citep{hunter06,hunter07}, and this is implemented in the statnet R
package \citep{handcock08,morris08,hunter2008ergm,statnet,ergm,krivitsky21}. However it is not
currently possible to estimate curved ERGMs using the EstimNetDirected
software \citep{stivala20}, and this is an area requiring further
work. In the absence of such a principled way of estimating the decay
parameters, an alternative to the heuristic (trial and error) approach
used here is to estimate many models with systematically varying
values of the $\lambda$ decay parameter for each relevant
``alternating'' model parameter, and use a grid search to find the
model with best fit.\footnote{This strategy was suggested by an
anonymous reviewer.} We applied this method to the Alon yeast
regulatory network model (\SI, Table~S2), using the
Mahalanobis distance between a vector of some of the observed network
summary statistics used for goodness-of-fit (degree distributions,
reciprocity, giant component size, global and average local clustering
coefficient), and the corresponding vectors for networks simulated
from the model, as the value to minimize. We used a two-dimensional
grid, varying the $\lambda$ value for AltOutStars as one dimension,
and the value of $\lambda$ for both AltTwoPaths and AltKTriangles
(these values should be the same, as described in \citet{snijders06})
as the other dimension. With both values varying from $1.5$ to $5.0$
in steps of $0.5$, we found the minimum Mahalanobis distance was
at $\lambda = 4.5$ for the AltOutStars parameter, and $\lambda = 1.5$
for the AltTwoPathsT and AltKTrianglesT parameters. The parameters
estimated for this model are not substantively different from those in
Table~\ref{tab:alon_yeast_lambda_ergm}. The values of $\lambda$ that we determined
heuristically (Table~\ref{tab:alon_yeast_lambda_ergm}) were at rank 15 (of
64) using this criterion.  The model with the default $\lambda = 2.0$
for all alternating statistic parameters, with subjectively poor
goodness-of-fit on the out-degree distribution, is at rank 48 (of 64).

As previously mentioned, the configurations available in an ERGM are
determined by the dependence assumptions: although there is a lot of
flexibility available in ERGM configurations, we cannot simply add
arbitrary configurations without regard for the underlying dependency
assumption \citep{koskinen20}. The least restrictive assumption used in
practice is the ``social circuit'' dependency assumption
\citep{snijders06,robins07,robins09,lusher13} used in this work, which
allows the use of the ``alternating'' configurations.

We also note that some recent work suggests that complex network
structure, including heavy-tailed degree distributions, closure
(clustering), large connected components, and short path lengths can
arise simply from thresholding normally distributed data to generate
the binary network \citep{cantwell20}.  Hence inferences from ERGM
modeling about network structure, just as with other techniques such
as comparison to ensembles of random graphs, could be consequences of
the way the binary network was constructed.

Valued ERGMs \citep{desmarais12,krivitsky12} may be used to avoid this
problem by removing the need to construct a binary network at all, and
working directly with the network with valued edges.
Parameter 
estimation for these models is even more computationally intensive than
for binary networks, and hence is so far impractical to use for
networks of the size considered here. Using new estimation techniques
to improve the scalability of parameter estimation for valued ERGMs
is another area requiring further research.

For the relatively small (on the order of one thousand nodes or fewer)
directed networks considered here, it is possible to do
simulation-based goodness-of-fit tests. However, it is possible to
estimate ERGM parameters for far larger (over one million nodes)
networks using the EstimNetDirected software, but it is not practical
to simulate such large networks from the model, and this is an area
requiring further work \citep{stivala20}.

One further limitation to consider is the execution time of the ERGM
technique. As discussed in the introductory sections, ERGM parameter
estimation is a computationally difficult problem. Although recent
advances allow the estimation in minutes of models that would have
taken hours, or been infeasible to estimate, with earlier methods, it
is still much more computationally difficult to do this than it is to
run conventional motif finding methods. The networks used here took
between three and 73 minutes to estimate, using multiple (up to 64) processor
cores in parallel. However motif finding with MFinder
\citep{kashtan04} in these networks takes only seconds, and with the
faster NetMODE method \citep{li12}, even less time, using only a
single processor core.

\section*{Conclusion}

We have re-examined the use of exponential random graph models
for analyzing biological networks, an application first introduced in the
bioinformatics literature by \citet{saul07}. Advances in ERGM
estimation methods since then have allowed more sophisticated models
to be estimated for more and larger networks than was possible at the
time, and they are now a more practical technique for making
inferences about structural hypotheses in biological networks, potentially solving
some of the problems inherent in conventional methods for
testing motif over-representation. By using an ERGM, all configurations in
the model are tested simultaneously, each conditional on all the
others, rather than having to test one at a time with the other
configurations fixed in a (more or less sophisticated, the choice of which is
critical to the results) null model.

The ERGM models of the Alon \textit{E.~coli} network presented here
are the first to retain the directed nature of the network and also
include terms for triangular structures. They confirm the result of
\citet{milo02} that path closure (feed-forward loop) is
over-represented, even when we include other, related, parameters in
the model.

We also presented the first ERGM models of a yeast regulatory network
retaining its inherently directed nature (rather than treating it as
undirected). We find statistically significant over-representation of
the transitive closure motif, just as \citet{milo02} did in
the same yeast regulatory network, using a simple randomization
test. 

The lack of the cyclic triangle (feedback loop) structure in the data,
however, is reproduced by models that do not contain any parameter
corresponding to this structure. This suggests that this structure is
not an ``anti-motif'', but rather that its lack is a consequence of
the structural features of the networks, specifically degree
distributions, two-paths, and transitive closure, that are included in
the models.

\section*{Acknowledgements}

Not applicable.

\section*{Funding}

This work was supported by Swiss National Science Foundation National
Research Programme 75 [grant number 167326]; and Melbourne
Bioinformatics at the University of Melbourne [grant number VR0261].

\section*{Abbreviations}

\begin{description}
\item[CDF:] Cumulative distribution function
\item[CUG:] Conditional uniform graph
\item[DMN:] Default mode network
\item[EE:] Equilibrium expectation
\item[EEG:] Electroencephalography
\item[ERGM:] Exponential random graph model
\item[GO:] Gene ontology
\item[HIPPIE:] Human integrated protein-protein interaction reference
\item[IFD:] Improved fixed density
\item[MAN:] Mutual, asymmetric, null
\item[MCMC:] Markov chain Monte Carlo
\item[PANTHER:] Protein analysis through evolutionary relationships
\item[PPI:] Protein-protein interaction
\item[SBM:] Stochastic block model
\end{description}

\section*{Availability of data and materials}
Source code, configuration files, and datasets are available from
\url{https://sites.google.com/site/alexdstivala/home/ergm_bionetworks}.


\section*{Competing interests}
The authors declare that they have no competing interests.


\section*{Authors' contributions}
AS conceived the work and conducted the analysis. AS and AL
interpreted the results. AS drafted the original manuscript and both
authors revised it. Both authors read and approved the final
manuscript.


\section*{Additional Files}
  \subsection*{Additional file 1 --- Supplementary tables and figures}
  Additional file 1 contains supplementary tables, including models of
  the \textit{E. coli} network estimated using statnet, and
  supplementary figures for additional goodness-of-fit plots.


\end{document}


\title{Testing biological network motif significance with exponential random graph models
  \\
  {\textit{Additional file 1}}}

\author{%
  Alex Stivala\affmark[1], Alessandro Lomi\affmark[1]\textsuperscript{,}\affmark[2] \\
  \affaddr{\affmark[1]Universit\`a della  Svizzera italiana, Via Giuseppe Buffi 13, 6900 Lugano, Switzerland} \\
  \affaddr{\affmark[2]The University of Exeter Business School, Rennes Drive, Exteter EX4 4PU,United Kingdom} \\
  \email{alexander.stivala@usi.ch}\\
  \email{alessandro.lomi@usi.ch}
}
\date{}

\maketitle

\setcounter{figure}{0}
\setcounter{table}{0}
\makeatletter
\renewcommand{\thefigure}{S\@arabic\c@figure} 
\renewcommand{\thetable}{S\@arabic\c@table} 
\makeatother

\section*{Supplementary tables}

\begin{table}[htb]
  \caption{Parameter estimates for the Alon \textit{E. coli}
    regulatory network, estimated using the ``stepping'' algorithm
    \citep{hummel12} in the statnet ergm package
    \citep{handcock08,morris08,hunter2008ergm,statnet,ergm,krivitsky21}.
    The decay parameter $\alpha$ for geometrically weighted dyad-wise
    shared partners (GWDSP) and geometrically weighted edge-wise
    shared partners (GWESP) OTP (``outgoing two path'', that is,
    transitive shared partner) is set to $\log(2.0)$, equivalent to
    the default value $\lambda = 2.0$ for AltTwoPaths and
    AltKTrianglesT in the EstimNetDirected software. }
  \label{tab:alon_ecoli_statnet_ergm}
\begin{tabular}{l*{2}{D{)}{)}{11)3}}}
\hline
Effect  & \multicolumn{1}{c}{Model 1} & \multicolumn{1}{c}{Model 2}\\
\hline
Edges  & -3.113 \; (0.077) ^{***} & -3.025 \; (0.066) ^{***}\\
GW in-degree ($\alpha = 2$)  & -3.954 \; (0.124) ^{***} & -3.818 \; (0.140) ^{***}\\
GW out-degree ($\alpha = 0$)  & 1.588 \; (0.184) ^{***} & 1.578 \; (0.190) ^{***}\\
GWDSP OTP ($\alpha = 0.693$)  & -0.493 \; (0.077) ^{***} & -0.502 \; (0.080) ^{***}\\
GWESP OTP ($\alpha = 0.693$)  & 2.378 \; (0.201) ^{***} & 2.382 \; (0.193) ^{***}\\
Nodematch self  &   & -0.271 \; (0.086) ^{**}\\
\hline
AIC  & 5445.00 & 5426.00\\
BIC  & 5495.00 & 5487.00\\
\hline
\end{tabular}
\floatfoot{*** $p < 0.001$; ** $p < 0.01$; * $p < 0.05$; $^{\boldsymbol{\cdot}} p < 0.1$.}

\end{table}

\begin{table}[h!]
  \caption{Parameter estimates for the Alon yeast
regulatory network, with the default $\lambda = 2$ for the
``alternating'' parameters.
    \label{tab:alon_yeast_ergm}}
    {\begin{tabular*}{\textwidth}{@{\extracolsep{\fill}}lrrr@{}}
      \toprule
      Effect  & Model 1 & Model 2 & Model 3\\
      \midrule
      Arc  & $\heavy{\underset{(-10.820, -10.335)}{-10.578}}$ & $\heavy{\underset{(-10.331, -9.828)}{-10.080}}$ & $\heavy{\underset{(-10.342, -9.844)}{-10.093}}$\\
      Sink  & $\light{\underset{(-3.407, 9.655)}{3.124}}$ & $\light{\underset{(-3.644, 9.721)}{3.039}}$ & $\light{\underset{(-3.637, 9.642)}{3.003}}$\\
      Source  & $\light{\underset{(-4.488, 10.295)}{2.903}}$ & $\light{\underset{(-6.715, 9.917)}{1.601}}$ & $\light{\underset{(-6.490, 10.389)}{1.949}}$\\
      Reciprocity  & --- & --- & $\light{\underset{(-26.307, 4.707)}{-10.800}}$\\
      AltInStars  & $\light{\underset{(-2.723, 3.416)}{0.347}}$ & $\light{\underset{(-2.804, 3.658)}{0.427}}$ & $\light{\underset{(-2.818, 3.490)}{0.336}}$\\
      AltOutStars  & $\light{\underset{(-0.278, 5.998)}{2.860}}$ & $\light{\underset{(-0.515, 5.685)}{2.585}}$ & $\light{\underset{(-0.537, 5.811)}{2.637}}$\\
      AltTwoPathsT  & --- & $\light{\underset{(-0.581, 0.360)}{-0.110}}$ & $\light{\underset{(-0.607, 0.310)}{-0.149}}$\\
      AltKTrianglesT  & $\heavy{\underset{(0.315, 2.139)}{1.227}}$ & $\light{\underset{(-0.773, 4.185)}{1.706}}$ & $\heavy{\underset{(0.011, 4.770)}{2.390}}$\\
      \bottomrule
    \end{tabular*}}

    \parbox{\textwidth}{
    Parameter estimates that are statistically significant at the 95\%
    level are shown in bold.  Model 3 is included for illustration,
    even though it shows poor convergence with respect to the
    Reciprocity parameter (t-ratio magnitude is greater than 0.3)
    }
\end{table}

\begin{table}[h!]
  \caption{Parameter estimates for the Alon yeast regulatory network,
    with self-edges included. The decay parameters $\lambda$ are set
    to the best value found by grid search in the network without
    self-edges, as described in the main text.
    \label{tab:alon_yeast_ergm_lambda_loops}}
{\begin{tabular*}{\textwidth}{@{\extracolsep{\fill}}lrr@{}}    
\toprule
Effect  & Model 1 & Model 2\\
\midrule
Arc  & $\heavy{\underset{(-7.606, -7.254)}{-7.430}}$ & $\heavy{\underset{(-7.702, -7.340)}{-7.521}}$\\
Loop  & --- & $\light{\underset{(-3.650, 7.091)}{1.720}}$\\
AltInStars   & $\light{\underset{(-1.498, 0.566)}{-0.466}}$ & $\light{\underset{(-1.414, 0.585)}{-0.414}}$\\
AltOutStars ($\lambda = 4.5$)  & $\heavy{\underset{(0.757, 1.243)}{1.000}}$ & $\heavy{\underset{(0.757, 1.252)}{1.005}}$\\
AltTwoPathsT ($\lambda = 1.5$)  & $\light{\underset{(-0.675, 0.068)}{-0.304}}$ & $\light{\underset{(-0.672, 0.084)}{-0.294}}$\\
AltKTrianglesT ($\lambda = 1.5$)  & $\heavy{\underset{(0.059, 3.709)}{1.884}}$ & $\heavy{\underset{(0.031, 3.704)}{1.868}}$\\
\bottomrule
\end{tabular*}}

    \parbox{\textwidth}{
    Parameter estimates that are statistically significant at the 95\%
    level are shown in bold.
    }
\end{table}

\clearpage

\section*{Supplementary figures}

\begin{figure}[ht!]
  \includegraphics[width=.9\textwidth]{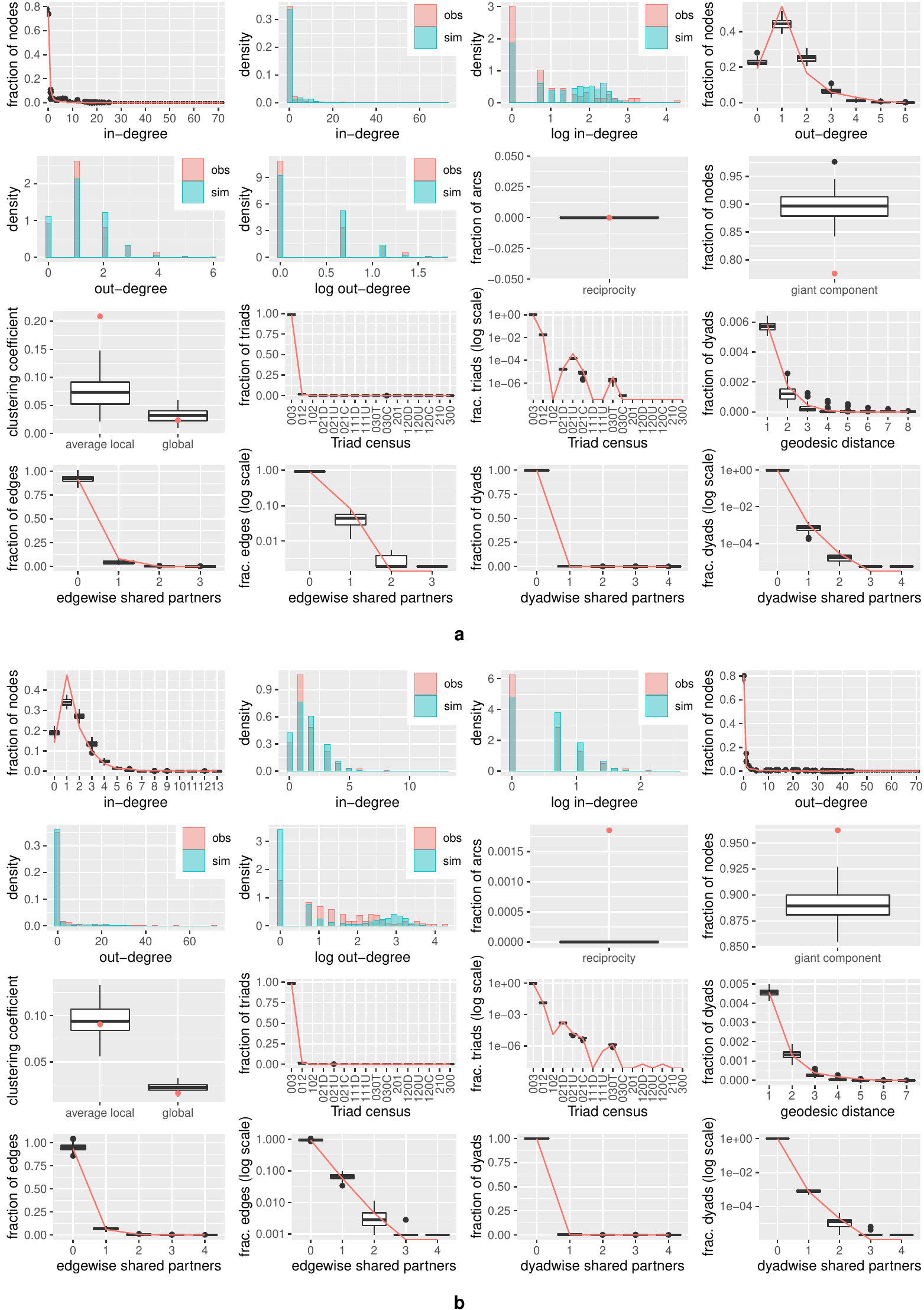}
  \caption{Goodness-of-fit plots for (\textbf{a}) the Alon \textit{E. coli}
    regulatory network Model 1 (Table~6 in the main text),
    and (\textbf{b}) the Alon yeast regulatory
    network Model 1 (Table~7 in the main text).
    The observed network statistics are plotted in red with the
    statistics of 100 simulated networks plotted as black boxplots,
    and blue on the histograms.
  \label{fig:alon_ecoli_yeast_gof}}
\end{figure}

\begin{figure}[ht!]
  \includegraphics[width=.9\textwidth]{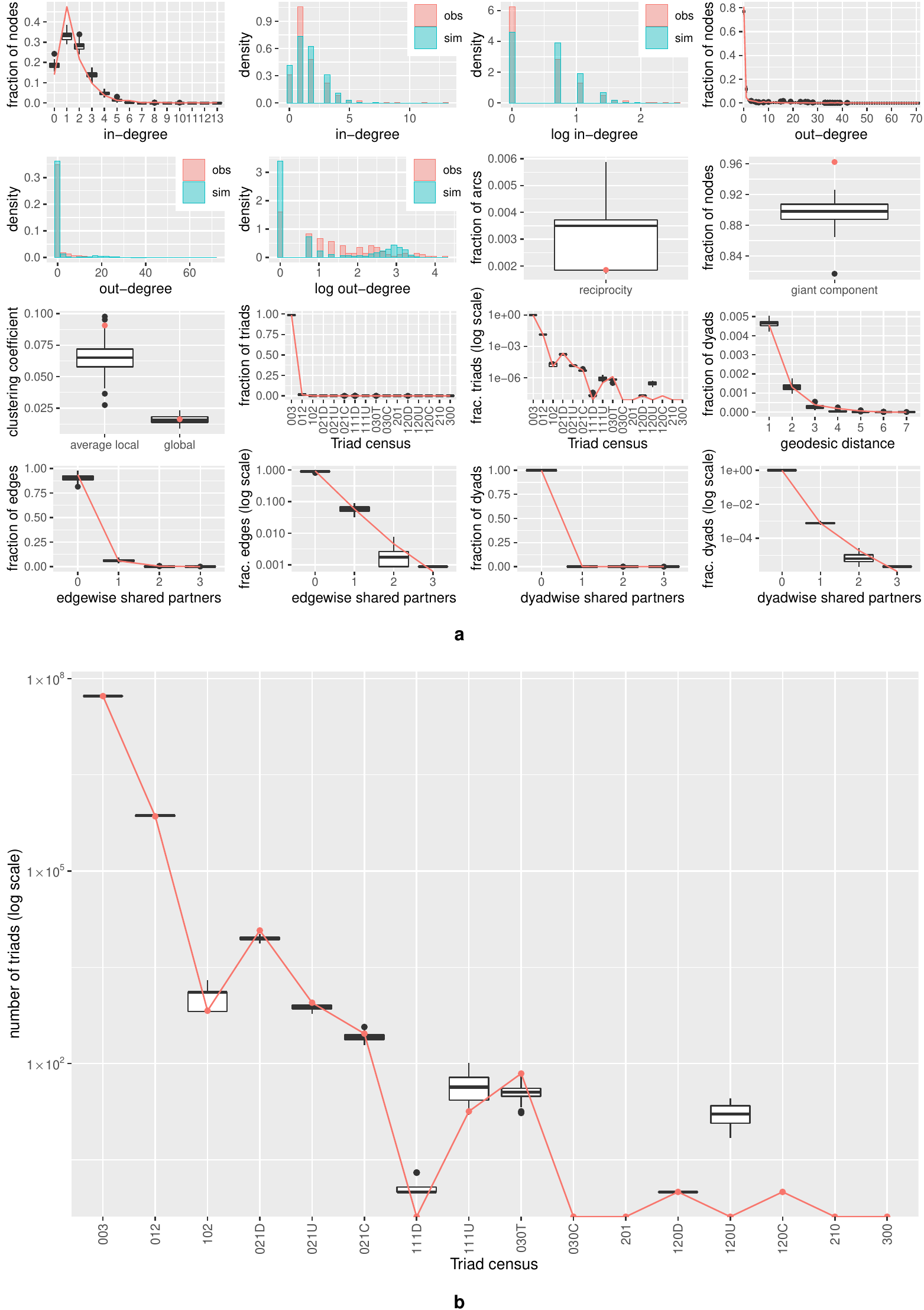}
  \caption{(\textbf{a}) Goodness-of-fit plots, and
    (\textbf{b}) triad census goodness-of-fit plot, for the Alon yeast regulatory
      network Model 2 (Table~7 in the main text).
    The observed network statistics are plotted in red with the
    statistics of 100 simulated networks plotted as black boxplots,
    and blue on the histograms.
  \label{fig:alon_yeast_lambda_model2_gof_plots}}
\end{figure}

\begin{figure}[ht!]
  \includegraphics[width=.9\textwidth]{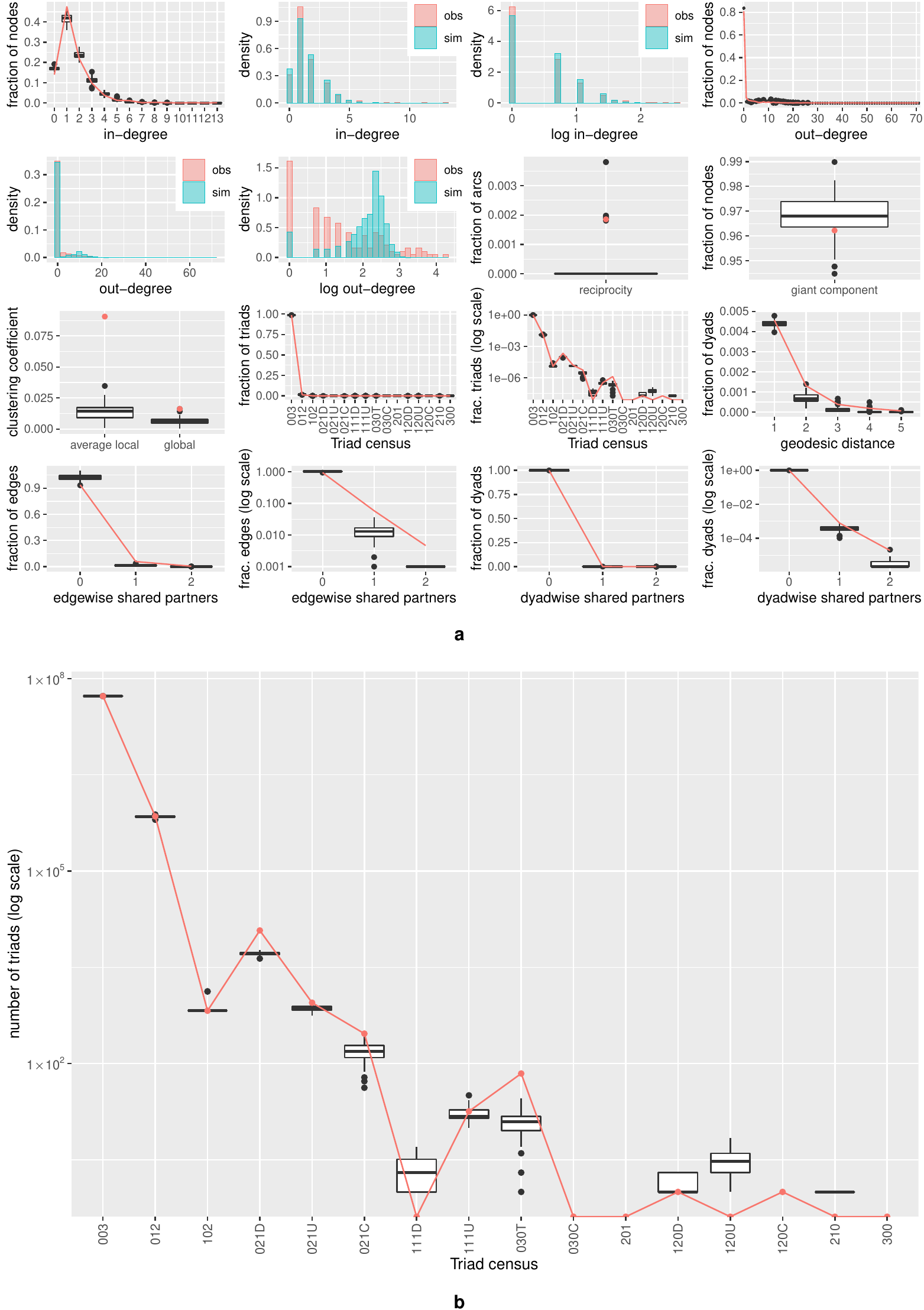}
  \caption{(\textbf{a}) Goodness-of-fit plots, and
    (\textbf{b}) triad census goodness-of-fit plot, for the Alon yeast regulatory
      network Model 2, with default decay parameter value $\lambda=2.0$ (Table~\ref{tab:alon_yeast_ergm}).
    The observed network statistics are plotted in red with the
    statistics of 100 simulated networks plotted as black boxplots,
    and blue on the histograms.
  \label{fig:alon_yeast_model2_gof_plots}}
\end{figure}

\begin{figure}
  \includegraphics[width=0.9\textwidth]{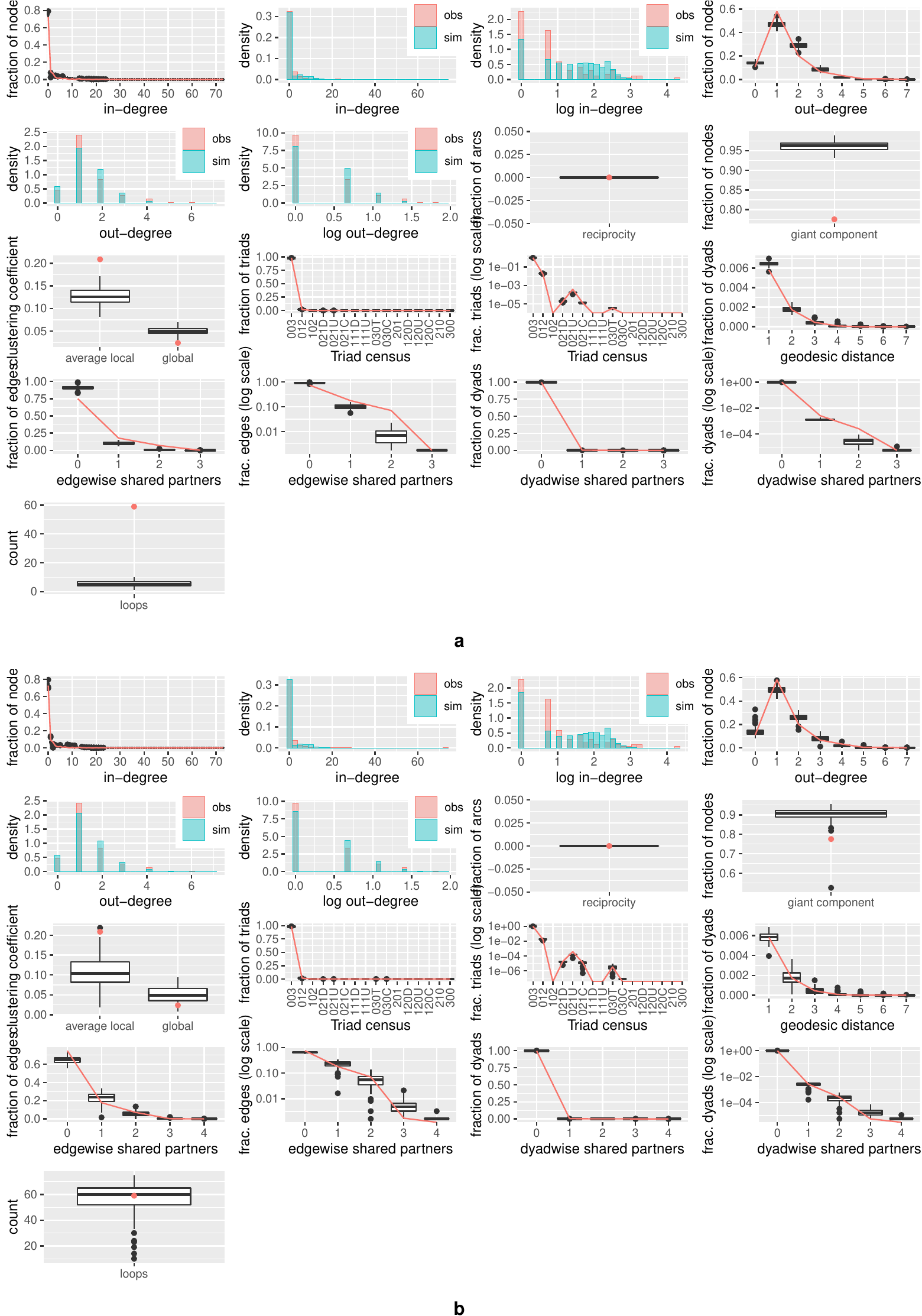}
  \caption{Goodness-of-fit plots for the Alon \textit{E. coli}
    regulatory network with self-edges (Models 3 and 4 in Table~6 in
    main text).  (\textbf{a}) Model 3, with no Loop parameter, and
    (\textbf{b}) Model 4, with the Loop parameter included.  }
\end{figure}

\begin{figure}
  \includegraphics[width=0.9\textwidth]{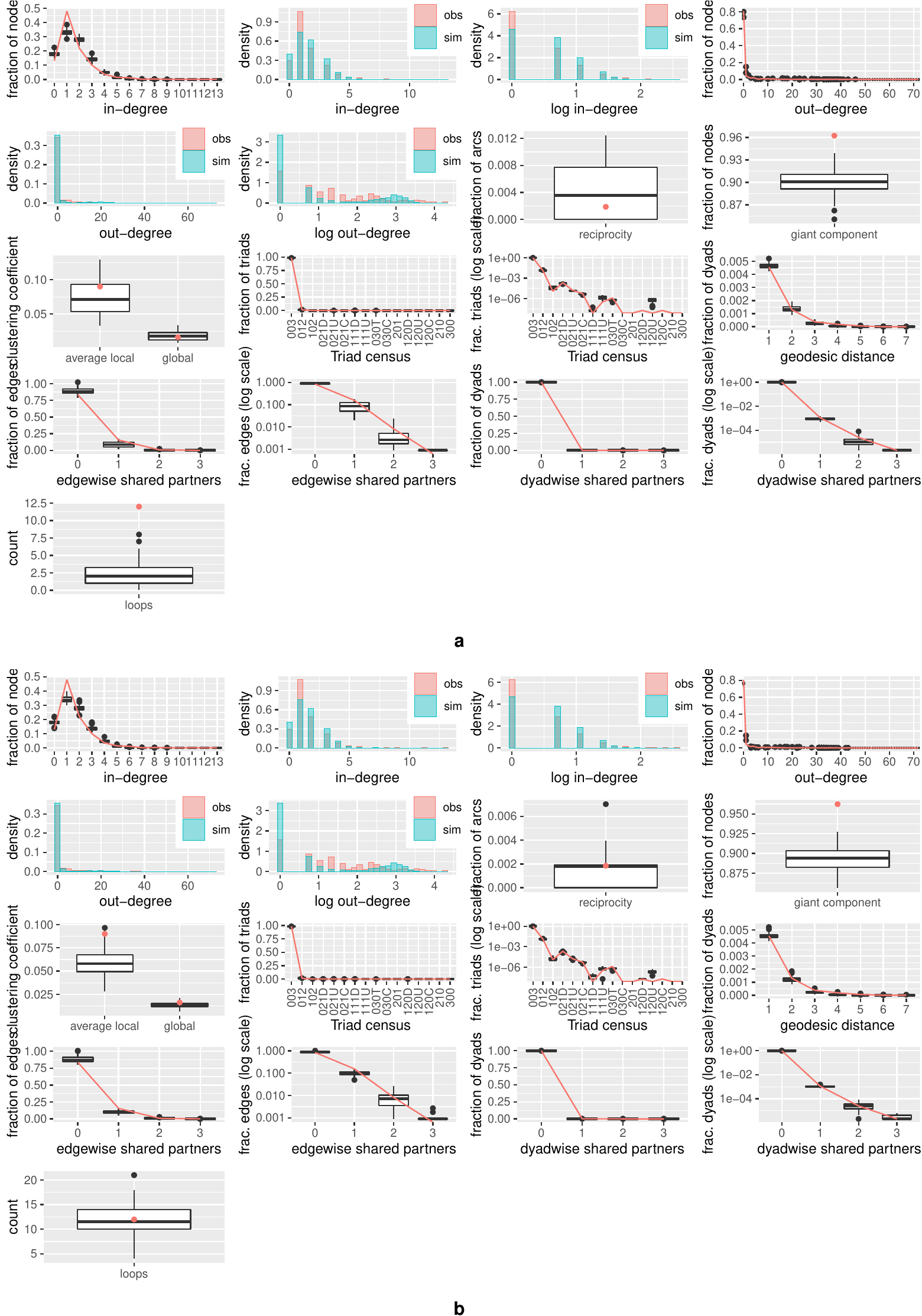}
  \caption{Goodness-of-fit plots for the Alon yeast regulatory
    regulatory network models with self-edges
    (Table~\ref{tab:alon_yeast_ergm_lambda_loops}).  (\textbf{a})
    Model 1, with no Loop parameter, and (\textbf{b}) Model 2, with
    the Loop parameter included.}
\end{figure}

\begin{figure}
  \includegraphics[scale=0.7,angle=270]{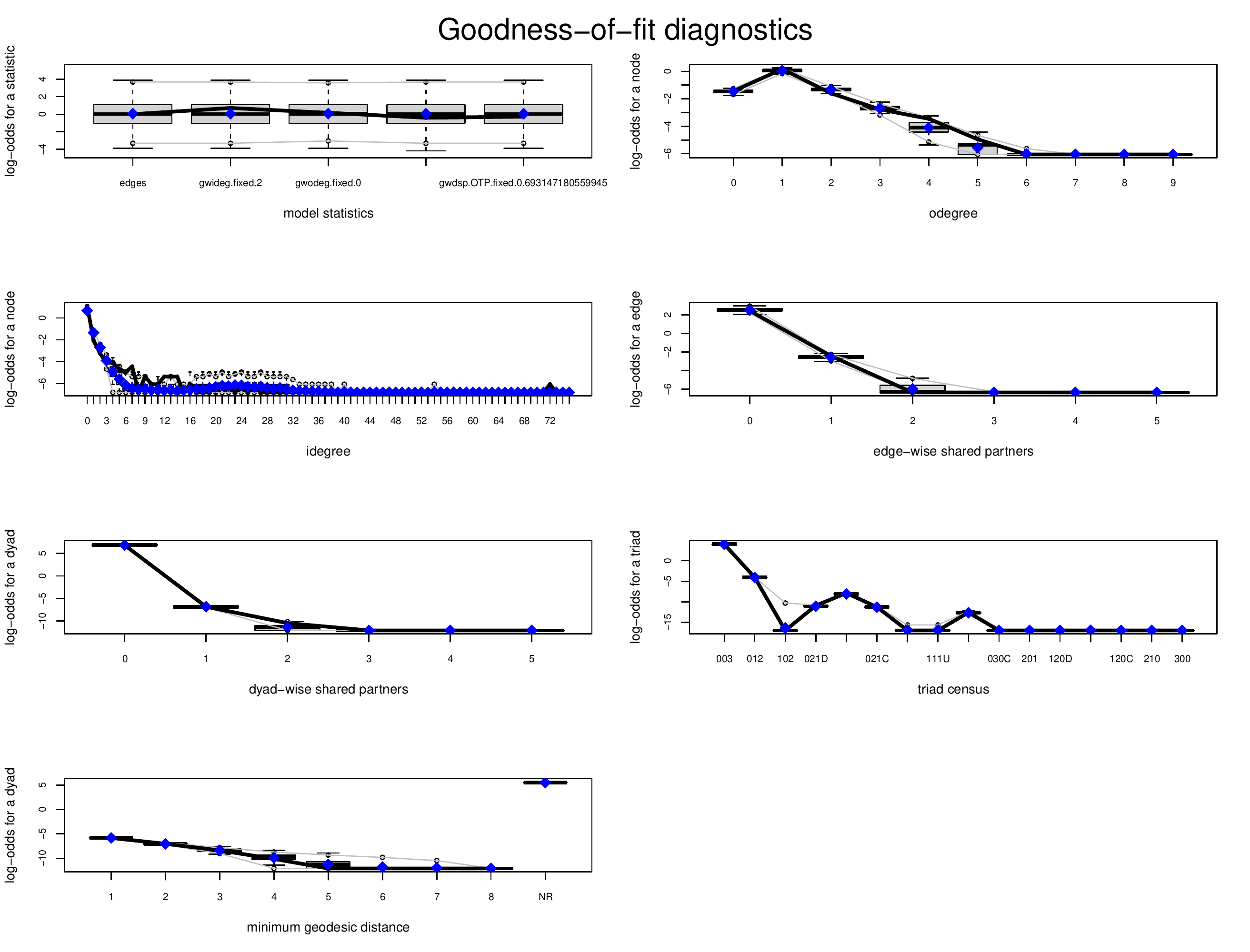}
  \caption{Statnet goodness-of-fit plots for the Alon \textit{E. coli}
    regulatory network, Model 1 (Table~\ref{tab:alon_ecoli_statnet_ergm}).}
\end{figure}

\begin{figure}
  \includegraphics[scale=0.7,angle=270]{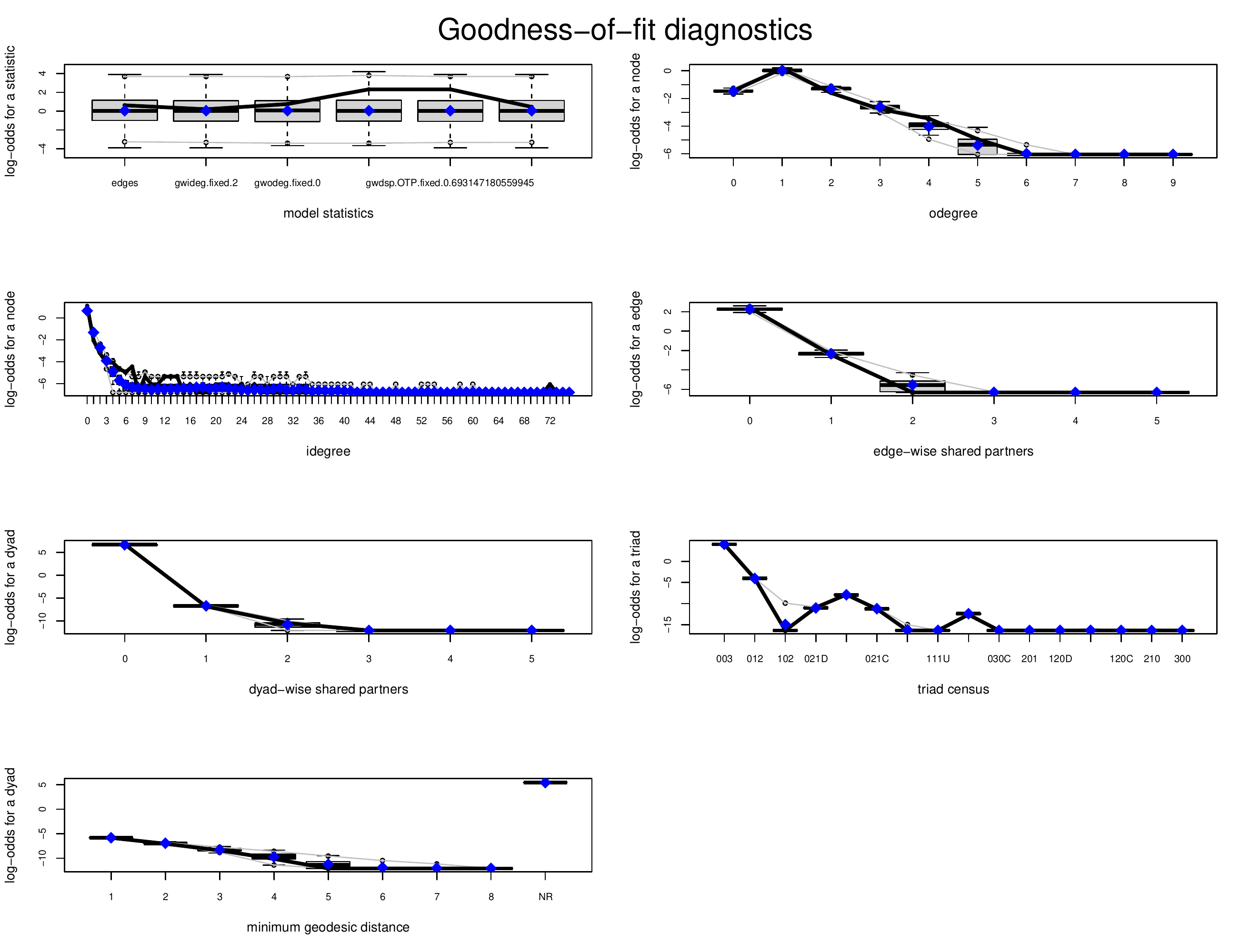}
  \caption{Statnet goodness-of-fit plots for the Alon \textit{E. coli}
    regulatory network, Model 2 (Table~\ref{tab:alon_ecoli_statnet_ergm}).}
\end{figure}

\clearpage
